\documentclass[preprint,pre,amsmath]{revtex4-2}

\usepackage{graphicx} 
\usepackage{bm}
\usepackage{epsfig}
\usepackage{latexsym}
\usepackage{amsmath}
\usepackage{nameref,hyperref}
\usepackage{cases,setspace,amsmath}
\usepackage{wasysym,amssymb}

\makeatletter
\let\old@makecaption=\@makecaption
\usepackage{subcaption}
\let\@makecaption=\old@makecaption
\makeatother
\usepackage{blindtext}

\usepackage{hyperref}

\def\av#1{{\langle  #1 \rangle}}
\def\be{\begin{equation}}
\def\ee{\end{equation}}
\def\bea{\begin{eqnarray}}
\def\eea{\end{eqnarray}}
\def\bsn{\begin{subnumcases}}
\def\esn{\end{subnumcases}}

\def\bml{\begin{mathletters}}
\def\eml{\end{mathletters}}
\def\bsn{\begin{subnumcases}}
\def\esn{\end{subnumcases}}

\def\nn{\nonumber}
\def\av#1{{\langle  #1 \rangle}}
\def\bav#1{{\big\langle #1 \big\rangle}}
\def\Bav#1{{\bigg\langle #1 \bigg\rangle}}

\begin{document}

\title{Dynamics of fixation probability in a population with fluctuating size}
\author{Kavita Jain} 
\email{jain@jncasr.ac.in}
\author{Hitesh Sumuni}
\email{hs23ms144@iiserkol.ac.in}
\affiliation{${}^\star$Theoretical Sciences Unit, \\Jawaharlal Nehru Centre for Advanced Scientific Research, \\Bangalore 560064, India \\
${}^\dagger$Department of Physical Sciences, Indian Institute of Science Education and Research Kolkata, Mohanpur, 741246, India}

\date{\today}

\begin{abstract}
In many biological processes, the size of a population changes stochastically with time, and recent work in the context of cancer and bacterial growth have focused on the situation when the mean population size grows exponentially. Here, motivated by the evolutionary process  of genetic hitchhiking in a selectively neutral population, we consider a model in which  
the mean size of the population increases linearly. 
We are interested in understanding how the fluctuations in the population size impact the first passage statistics, and study the fixation probability that a mutant reaches 
frequency one by a given time in a population whose size follows a conditional Wright-Fisher process. 
We find that at sufficiently short and long times, the fixation probability  can be approximated by a model that ignores temporal correlations between the inverse of the population size, but at intermediate times, it is significantly smaller than that obtained by neglecting the correlations. Our analytical and numerical study of the correlation functions show that the conditional Wright-Fisher process of interest is neither a stationary nor a Gaussian process; we also find that the variance of the inverse population size initially increases linearly with time $t$ and then decreases as $t^{-2}$ at intermediate times followed by an exponential decay at longer times. Our work emphasizes the importance of temporal correlations in populations with fluctuating size that are often ignored in population-genetic studies of biological evolution.  
\end{abstract}
 
\maketitle

\clearpage
\section{Introduction}

The remarkable genetic and phenotypic diversity on earth is a product of biological evolution 
\cite{Darwin:1859} which 
 is a complex process involving several fundamental processes, {\it viz.}, mutation, selection and migration, and various sources of noise \cite{Charlesworth:2010}. Often these processes act simultaneously and their magnitude and/or direction varies with time; for example, the selection pressure or size of a population may change due to seasonal cycles. While much theoretical work has been done assuming that the variation in selection \cite{Ohta:1968,Waxman:2011,Uecker:2011,Devi:2020,Assaf:2008,Kaushik:2021,Jain:2022,Kaushik:2023} or population size \cite{Nei:1975,Maruyama:1985c,Williamson:2005,Waxman:2012,Nakamura:2018,Kaushik:2025} is deterministic, in biologically realistic situations, these parameters are random variables. Several recent work have focused on addressing this aspect and assume that the population parameters follow a stochastic process such as a Gaussian process \cite{Takahata:1975,Gillespie:1991,HuertaSanchez:2008}; two-state process with the distribution of the switching times between the states to be an exponential (random telegraph process) \cite{Sjodin:2005,Schaper:2012,Zivkovic:2015,Asker:2025,Jain:2025} or a power law \cite{Meyer:2024}; and density-dependent birth-death process  \cite{Lambert:2006,Parsons:2007a,Parsons:2007b,Parsons:2010,Engen:2009,Czuppon:2018}.

In the last decade, motivated by the rapid growth of cancer or bacterial cells,  the statistical properties of the mutant subpopulation in a population whose size fluctuates in time such that its mean size grows exponentially have been investigated \cite{Kendall:1960,Durrett:2013,Kessler:2013,Durrett:2015,Ohtsuki:2017,Cheek:2018,Gunnarsson:2021,Nicholson:2023}.  In these studies, the growing population size is modeled as a birth-death process and the mean number of mutants at a genomic site has been analyzed in the framework of a branching process.  Here, we are interested in the evolutionary dynamics in a population whose mean size grows {\it linearly} with time due to genetic hitchhiking in a selectively neutral population \cite{Mafessoni:2015,Moinet:2022}, as described below. 

\begin{figure}[t]
     \centering
    \begin{subfigure}{0.49\textwidth}
         \centering
         \includegraphics[width=1\textwidth]{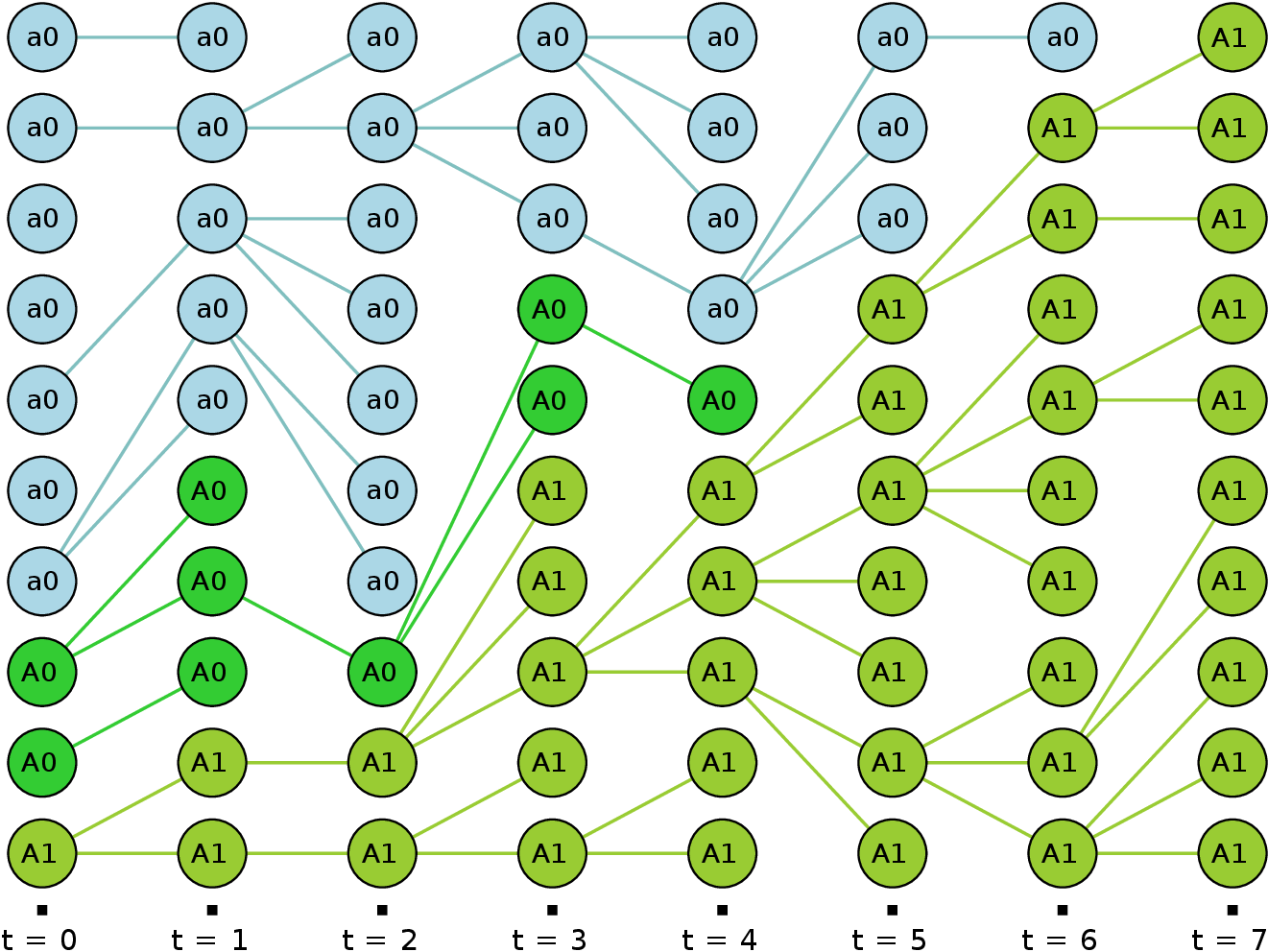}
                  \caption{}
         \label{fig_sweep}
     \end{subfigure}
     \begin{subfigure}{0.49\textwidth}
         \centering
         \includegraphics[width=1\textwidth]{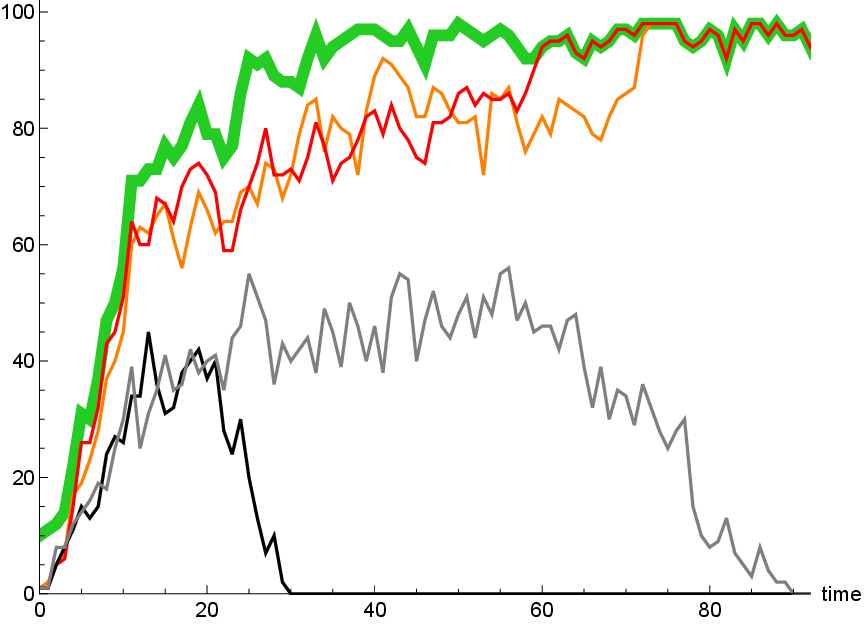}
         \caption{}
               \label{fig_trajecs}
     \end{subfigure}
     \caption{(a) Schematic description of the model depicting Wright-Fisher dynamics in a population of total constant size $N=10$. At $t=0$, $N_0=3$ individuals carry $A$ allele at the first site and at $t=7$, $A$ fixes in the population. Starting with a single allele $1$ ($n_1=1$) in the subpopulation of $A$s, the  number of $1$s are tracked until the entire population of $A$s is either $A0$ or $A1$ type; note that the allele $1$ fixes in the $A$ subpopulation at $t=5$. 
The effect of the dynamics of $A$ allele on the dynamics at the second site can be incorporated as fluctuating population size as described in the text. (b) Sample trajectories of the $A1$ subpopulation (thin lines) for a given trajectory of the $A$ allele that eventually fixes (thick line) for $n_1=1, N_0=10, N=100$.}
       \label{fig_1}
\end{figure}

We consider a finite population of binary sequences with two linked sites where the first site is, in general, under selection and the second one is neutral (that is, the wildtype and mutant type/allele are equally fit). 
During the evolutionary process, as shown in Fig.~\ref{fig_sweep}, if the mutant allele $A$ at the first site escapes loss due to stochastic fluctuations and  fixes in the population (that is, it reaches a frequency one),  
then the frequency of neutral allele $1$ initially attached to it also rises and allele $1$ thus hitchhikes  to fixation \cite{Smith:1974}. This phenomenon has been shown to result in a decrease in the linked neutral genetic diversity in general models as well as sequence data  \cite{Charlesworth:2021},  
and was proposed as a mechanism to explain the Lewontin's paradox \cite{Lewontin:1974,Buffalo:2021} which refers to the observation that the neutral genetic diversity seen in natural populations is much smaller than that predicted by the neutral theory \cite{Kimura:1983}. As Fig.~\ref{fig_sweep} suggests, the dynamics at the second site in the subpopulation of $A$s can be modeled by that of a neutral population of $0$s and $1$s whose total size is changing with time. If allele $A$ is strongly beneficial, the variation in population size can be captured by a deterministic model in which the population size grows exponentially  \cite{Kaushik:2025} (in related contexts, Gaussian fluctuations about the deterministically growing population have also been considered \cite{Terhorst:2015,Blomberg:2020,Devi:2023}), otherwise the total population size fluctuates and, in our model of interest, its mean size increases linearly at short times.

In this article, we focus on understanding how the time-dependent fixation probability of a mutant is affected due to stochastic changes in the population size. In Sec.~\ref{sec_model}, we define the model and describe the relevant Fokker-Planck equations. The known explicit solution of these equations when the population size is constant is discussed in Sec.~\ref{FPEcst} and a formal solution of the model when the population size fluctuates is given in Sec.~\ref{FPEfluc}.  Due to temporal correlations in the population size, the model of interest does  not seem to be exactly solvable, and therefore we first study the fixation probability ignoring these correlations in Sec.~\ref{UncorrM}. Then in Sec.~\ref{CorrM}, we study a two-point correlation function analytically and higher cumulants numerically which show that the process that describes the changing population size is not a stationary Gaussian process, and discuss how correlations affect the fixation probability. We close the article with a discussion of related models and future directions in Sec.~\ref{sec_disc}. 

\section{Model}
\label{sec_model}

We consider a finite population of binary sequences of length two where the allelic state at the first (second) site can be either $a$ or $A$ ($0$ or $1$), no mutations between the two alleles at either site occur during the dynamics and all the four sequence configurations are equally fit (but, see Appendix~\ref{app_genl} for a general model). At the first site, the dynamics follow a discrete time, neutral Wright-Fisher process for a population of {\it constant} size $N$; in this process, irrespective of the state at the second site, the number $n'_A$ of mutant allele $A$ in the current generation is binomially distributed with mean equal to the number $n_A$ of $A$s in the previous generation. Thus the transition probability of the Markov chain that describes this process is given by 
\be
T(n'_A|n_A)={N \choose n'_A} \left(\frac{n_A}{N}\right)^{n'_A} \left(1-\frac{n_A}{N}\right)^{N-n'_A} \label{neuBin}
\ee

We are interested in the {\it conditional} process in which, starting from $2 \le N_0 < N$,  allele $A$ eventually fixes (henceforth referred to as the $A^*$ process). 
In this subpopulation of $A$s, we track the number of $1$s at the second site whose dynamics also follow the neutral Wright-Fisher process but with changing population size due to the dynamics of allele $A$. Thus, in the current generation, the number of $1$s in a  subpopulation of size $n'_A$ are binomially distributed with mean given by $n'_A \times \frac{n_1}{n_A}$ where $n_1$ is the number of $1$s in the  previous generation. 
A schematic illustration of the model and examples of stochastic trajectories of allele $1$ while the $A$ allele is proceeding to fixation are shown in Fig.~\ref{fig_1}. Note that the fixation of allele $1$ can occur when the population size in the $A^*$ process is smaller than $N$. 
We simulated the model described above and obtained the quantities of interest by averaging over  $10^3-10^4$ fixations of the allele $1$ for each trajectory of the $A$ allele in $10^3-10^4$ independent runs of the $A^*$ process. 

In continuous time, the distribution $P(p, t; p_0, 0)$ of the frequency $p=\frac{n_A}{N}$ of $A$s at time $t$, starting with frequency $p_0=\frac{N_0}{N}$ can be described by the following forward Fokker-Planck equation (FPE) \cite{Kimura:1964,Ewens:2004}:
\bea
\frac{\partial P}{\partial t} &=& \frac{1}{2 N} \frac{\partial^2}{\partial p^2} \left( p(1-p) P \right) \label{FPEP}
\eea
This can be seen by noting that due to Eq.~(\ref{neuBin}), the mean and variance of the change in the frequency $\delta p$ in one generation  
are, respectively, zero and $\frac{p(1-p)}{N}$, and  that the higher moments in $\delta p$ vanish in the scaling limits $t \to \infty, N \to \infty, \frac{t}{N}$ finite. Then from Bayes' theorem, the above unconditional distribution $P(p,t; p_0,0)$ can be related  to the distribution $P^*(p,t; p_0,0)$ conditioned on the fixation of $A$ as
\be
P^*(p,t; p_0,0)=\frac{p}{p_0} ~ P(p,t; p_0,0) \label{Bayes}
\ee
on using that the eventual fixation probability of $A$ is equal to its initial frequency $p_0$  \cite{Kimura:1964,Ewens:2004} (also see Eq.~\ref{Pi1} below). Finally, for a given stochastic trajectory of $A^*$ process with frequency $p(t)$, as for Eq.~(\ref{FPEP}), the distribution $X(x, t; x_0,0)$ of the frequency $x=\frac{n_1}{n_A}$ of $1$s with initial frequency $x_0$ obeys the following FPE \cite{Risken:1996,Ewens:2004,Kaushik:2025}, 
\bea
\frac{\partial X}{\partial t} &=& \frac{1}{2 N p(t)} \frac{\partial^2}{\partial x^2} \left( x(1-x) X\right) \label{FPEF}
\eea


 \section{Distributions for constant population size}
 \label{FPEcst}
 
We first describe the known results for the model with constant population size that are pertinent to the discussion here. At $t \to \infty$, the subpopulation comprising of $A$s at the first site either goes extinct or gets fixed, but at finite times, the population can either exist in one of these two absorbing states or have a finite frequency $0 < p <1$. The solution of Eq.~(\ref{FPEP}) can then be written as \cite{McKane:2007}
\be
P(p,t;p_0,0)= \delta(p) \Pi_0(t; p_0)+ \delta(1-p) \Pi_1(t; p_0)+ \Theta(p) \Theta(1-p) f(p,t; p_0,0) \label{pieces}
\ee
Here, $\Pi_0(t; p_0)$ and $\Pi_1(t; p_0)$ are, respectively, the extinction and fixation probability of $A$ by time $t$, while $f(p,t; p_0,0)$ is the distribution of frequency $0 < p < 1$ at time $t$. 

Due to the conservation of probability for $0 \leq p \leq 1$, it follows from Eq.~(\ref{FPEP}) that 
 the probability current, $J(p,t)=-\frac{1}{2 N}  \frac{\partial}{\partial p} \left( p(1-p) P \right)$ vanishes at $p=0$ and $1$; furthermore, the distribution $P$ is normalizable if it diverges weakly enough close to the boundaries. Thus we have   \cite{McKane:2007}
 \be
 \textrm{Lim}_{p \to 0, 1} J(p, t)=0, \;\; \textrm{Lim}_{p \to 0, 1} ~p (1-p) P(p,t)=0 \label{bcP}
 \ee
 for all $t$. Subject to these boundary conditions, Eq.~(\ref{FPEP}) can be solved using the eigenfunction expansion method on noting that the Jacobi polynomials $P_n^{(\alpha,\beta)}(p)$ obey (see (18.8.1) of \cite{DLMF}), 
\be
\frac{\partial^2}{\partial p^2} [p (1-p) P^{(1,1)}_{m-1}(1-2p)]=-m (m+1) P^{(1,1)}_{m-1}(1-2 p) ~,~m=1, 2, ...
\ee
For the initial condition, $P(p,0; p_0,0)=\delta(p-p_0)$, one then obtains \cite{Kimura:1955a,McKane:2007}
\bea
\Pi_0(t; p_0) &=& 1-p_0 -p_0 (1-p_0)  \sum_{n=1}^\infty   \frac{2 n+1}{n} P^{(1,1)}_{n-1}(1-2 p_0) e^{-\Lambda_{n} t} \label{Pi0} \\
\Pi_1(t; p_0) &=& p_0+p_0(1-p_0)  \sum_{n=1}^\infty   \frac{(-1)^n (2 n+1)}{n} P^{(1,1)}_{n-1}(1-2 p_0) e^{-\Lambda_{n} t} \label{Pi1}\\
f(p,t; p_0,0) &=&  p_0 (1-p_0)\sum_{n=1}^\infty   \frac{(2 n+1) (n+1)}{n} P^{(1,1)}_{n-1}(1-2 p_0) P^{(1,1)}_{n-1}(1-2 p) e^{-\Lambda_{n} t} \label{fpt} 
\eea
where the eigenvalue, 
\be
\Lambda_n=\frac{n (n+1)}{2 N}~,~n=1, 2, ... \label{eval}
\ee
From Eq.~(\ref{Pi0}) and Eq.~(\ref{Pi1}), we note that the probability of eventual extinction and fixation are, respectively,  given by $1-p_0$ and $p_0$.

For numerical purposes, it is convenient to find the absorption probabilities using a backward Fokker-Planck equation given by \cite{Kimura:1964,Ewens:2004}, 
\bea
\frac{\partial \Pi}{\partial t} &=& \frac{p_0 (1-p_0)}{2 N} \frac{\partial^2 \Pi}{\partial p_0^2} \label{bFPE}
\eea
with boundary conditions, $\Pi_0(t; 0)=1, \Pi_0(t; 1)=0$ and $\Pi_1(t; 0)=0, \Pi_1(t; 1)=1$, and initial condition $\Pi_0(0; p_0)=\Pi_1(0; p_0)=0$ for $0 < p_0 < 1$. 

\begin{figure}[t]
     \centering
    \begin{subfigure}{0.49\textwidth}
         \centering
         \includegraphics[width=1.\textwidth]{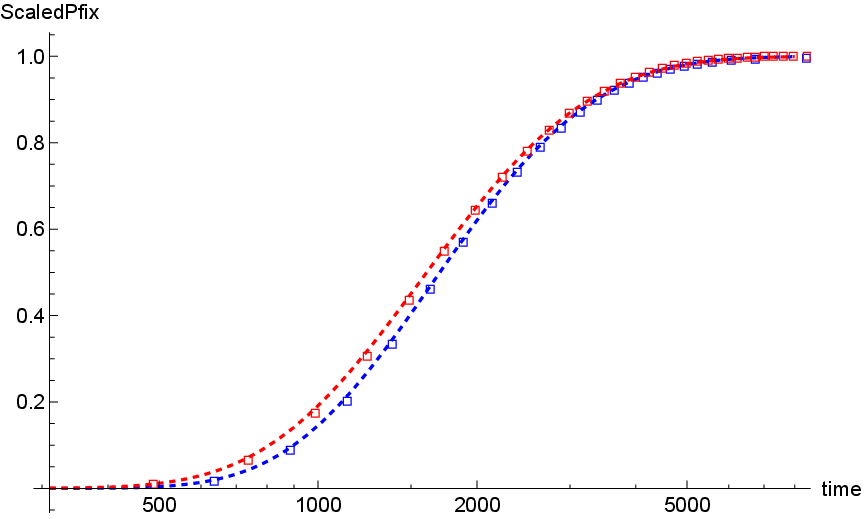}
                 \caption{}
         \label{fig_fixN}
     \end{subfigure}
     \begin{subfigure}{0.49\textwidth}
         \centering
          \includegraphics[width=1.\textwidth]{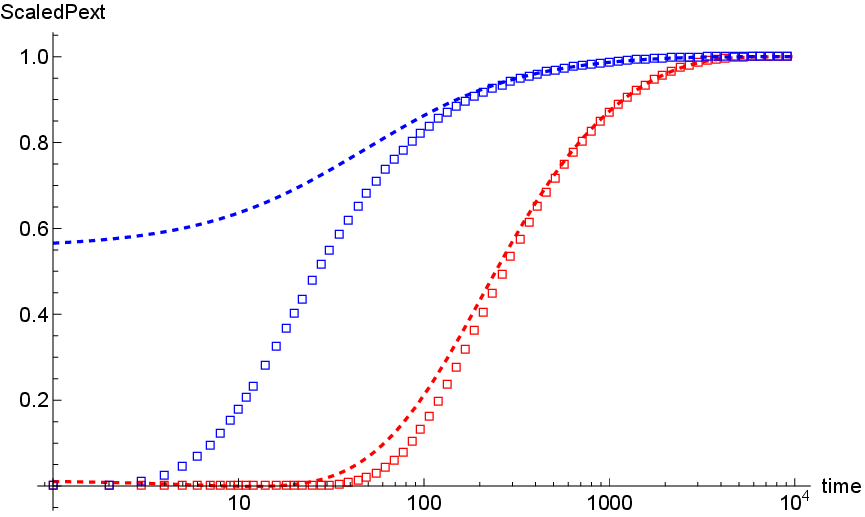}
         \caption{}
               \label{fig_extN}
     \end{subfigure}
     \caption{(a) Scaled fixation probability, $\frac{\Pi_1(t; p_0)}{p_0}$ and (b) scaled extinction probability, $\frac{\Pi_0(t; p_0)}{1-p_0}$ of mutant allele $A$ in a population of constant size $N=1000$, and $N p_0=N_0=10$ (blue) and $100$ (red). The points show the simulation results and the dashed lines are obtained by solving Eq.~(\ref{bFPE}) numerically.} 
        \label{fig_fixext}
\end{figure}

The simulation results for the dynamics of fixation and extinction probability of $N_0$ number of $A$s in a population of constant size $N$ are shown in Fig.~\ref{fig_fixN} and  Fig.~\ref{fig_extN}, respectively. As intuitively expected, more the initial number of $A$s, higher the chance of fixation and lower the chance of extinction. But while the fixation probability is seen to depend weakly on the initial number of mutants, the extinction probability depends on $N_0$ and remains negligible on a time scale that increases linearly with $N_0$. This is because (all) the descendants of all the $A$ individuals that are  initially present must die for the extinction of allele $A$ to occur, but $A$ gets fixed if the lineage of any one of the initial $A$s survives.

Figure~\ref{fig_fixext} also shows a comparison between the simulation data and the results obtained by numerically solving Eq.~(\ref{bFPE}) with appropriate boundary conditions. 
We note that while they agree for the fixation probability, there is a strong disagreement for the extinction probability for smaller $N_0$ at short times  - this is because the extinction probability is non-negligible on times of order $N_0$ but, as explained below Eq.~\ref{FPEP}, the FPE is valid for large times.  An analytical understanding of the relevant time scales predicted by Eq.~\ref{Pi0} and Eq.~\ref{Pi1} is obtained in Appendix~\ref{app_HM}.

\section{Fixation probability for fluctuating population size}
\label{FPEfluc}

The probability distribution described by Eq.~(\ref{FPEF}) for changing population size can be obtained in a manner analogous to that for the constant-sized population if one defines the time variable to be $T=\int_0^t \frac{dt'}{p(t')}$. Then, using Eq.~(\ref{Pi1}), the fixation probability, $\Phi_1(t; x_0)$ of the mutant allele $1$, starting from frequency $x_0$, can be written as 
\bea
\Phi_{1}(t; x_0)=x_0+x_0(1-x_0)  \sum_{n=1}^\infty   \frac{(-1)^n (2 n+1)}{n} P^{(1,1)}_{n-1}(1-2 x_0) e^{-\Lambda_{n} \int_0^t \frac{dt'}{p(t')}} \label{Pi1F}
\eea
At $t \to \infty$, as $p \to 1$, it follows from the above equation that even for changing population size, the eventual fixation probability is given by the initial frequency $x_0$.   

On averaging over the fluctuating population size, we obtain $\av{\Phi_1}^*$ where the starred angular brackets denote the  average with respect to the frequency distribution in the $A^*$ process. Equation (\ref{Pi1F}) shows that for this purpose, we  require \cite{Kubo:1962}
\bea
\bav{e^{-\Lambda_n\int_0^t \frac{dt'}{p(t')}}}^* &=& \exp \left[\sum_{m=1}^\infty (-\Lambda_n)^m \int_0^t dt_1 \int_0^{t_1} dt_2 ... \int_0^{t_{m-1}} dt_m  \Bav{\frac{1}{p(t_1)} ... \frac{1}{p(t_m)}}^*_c \right]  \label{cumexpn} \\
&=& \exp \left[ -\Lambda_n \int_0^t dt_1 \Bav{\frac{1}{p(t_1)}}^* +\Lambda_n^2 \int_0^t dt_1 \int_0^{t_1} dt_2 \Bav{\frac{1}{p(t_1) p(t_2)}}^*_c-... \right] \label{cumexpn2}
\eea
where the subscript $c$ denotes the cumulant. If the $A^*$ process for the inverse frequencies is a stationary Gaussian process, the above expression simplifies considerably (see, for e.g., Eq.~(3.75) of \cite{Risken:1996}). But as shown below, this process is neither stationary nor a Gaussian process, and it does not appear possible to obtain an exact expression for the fixation probability in the full model. 
Therefore, in Sec.~\ref{UncorrM}, we first consider a model that ignores all the temporal correlations, and then study the nature of the correlations and their effect on fixation probability in Sec.~\ref{CorrM}.

\begin{figure}[t]
     \centering
    \begin{subfigure}{0.49\textwidth}
         \centering
         \includegraphics[width=1.0\textwidth]{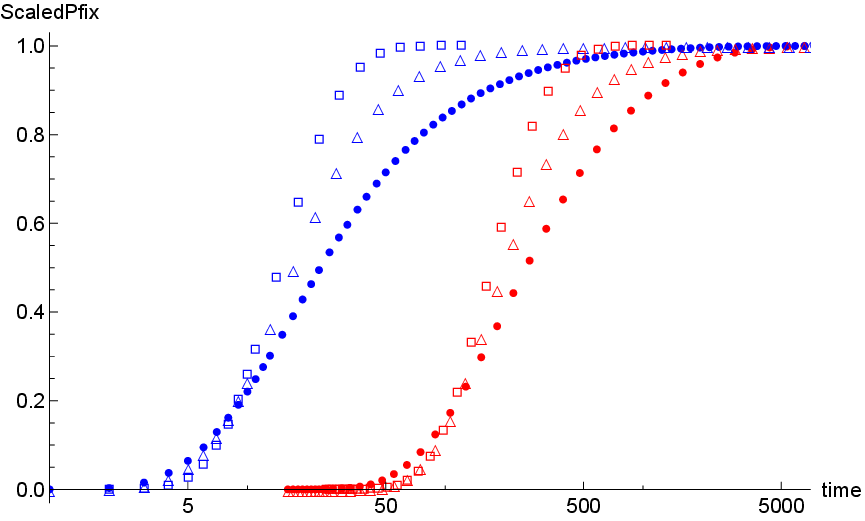}
         \caption{}
         \label{fig_fullun}
     \end{subfigure}
     \begin{subfigure}{0.49\textwidth}
         \centering
         \includegraphics[width=1.0\textwidth]{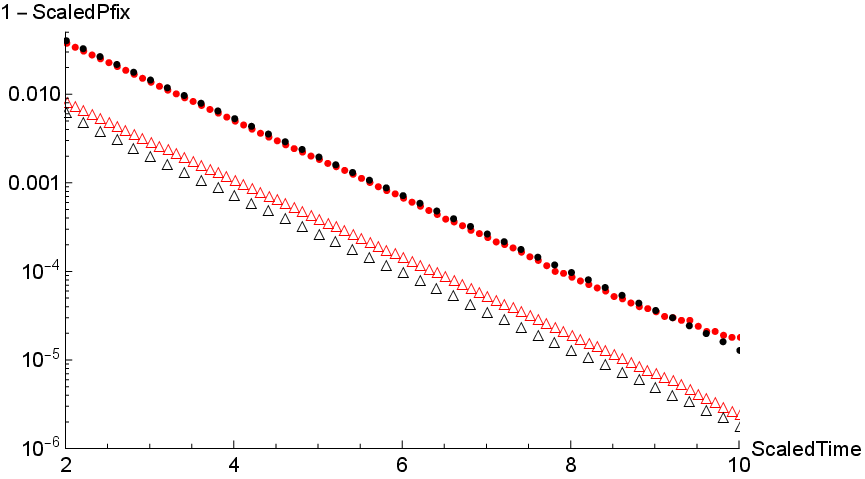}
         \caption{}
               \label{fig_appro}
     \end{subfigure}
     \caption{(a) Fixation probability of a single mutant allele $1$ in a subpopulation of $A$s with initial size $N_0=10$ (blue) and $100$ (red) divided by the initial frequency, $x_0=N_0^{-1}$. The simulation data for the full model ($\bullet$) and uncorrelated model ($\triangle$) for $N=1000$, and when population size remains constant at the initial size ($\square$) are  shown.  (b) Figure shows the complementary scaled fixation probability as a function of scaled time, $\frac{t}{N}$ for $N_0=100, N=1000$ (red) and $N_0=50, N=500$ (black) for full model ($\bullet$) and uncorrelated model ($\triangle$).}
       \label{fig_pfix}
\end{figure}

 Figure~\ref{fig_fullun} shows the simulation results for the fixation probability of a single mutant $1$ when the population size is changing in the full model defined in Sec.~\ref{sec_model} and in the uncorrelated model described in Sec.~\ref{UncorrM}, and when the population size remains fixed at the initial size $N_0$. We note that until time of order $N_0$, the fixation probability is almost the same in the three models. But at larger times $t \gg N_0$, the mutant allele $1$ is more likely to fix when the population size remains constant at $N_0$ than when it is increasing as it is harder to fix in a larger population. Furthermore, as displayed in Fig.~\ref{fig_appro}, at times of order $N$, the fixation probability in the full model can be approximated by that in the uncorrelated model. However, as Fig.~\ref{fig_fullun} shows, at intermediate times, fixation is much more likely in the uncorrelated model than in the full model; see Sec.~\ref{sec_fullfix} for a discussion. 
 
\section{Uncorrelated population size}
\label{UncorrM}

In this section, we assume that the inverse frequencies in the $A^*$ process are uncorrelated in time. Then Eq.~(\ref{cumexpn}) gives 
\bea
\av{e^{-\Lambda_n\int_0^t \frac{dt'}{p(t')}}}^* &\stackrel{uncorr}{=}& e^{-\Lambda_n \int_0^t dt' \bav{\frac{1}{p(t')}}^*}=e^{-\Lambda_n \int_0^t \frac{dt'}{p^*_{hm}(t')}} \label{uncorr1}
\eea
where, $p^*_{hm}(t)=[\bav{\frac{1}{p(t)}}^*]^{-1}$ is the  time-dependent, conditional harmonic mean frequency, and is  given by 
\be
p^*_{hm}(t)=\frac{p_0}{1-\Pi_0(t; p_0)}\label{phmdef}
\ee
as shown below [see Eq.~(\ref{cHM})]. 

 \subsection{Conditional mean and harmonic mean}

We now discuss how the arithmetic mean and harmonic mean of frequency $p$ change with time. Evidently, these quantities are identical in the full model and the uncorrelated model. We first consider the (arithmetic) mean frequency in the $A^*$ process which is given by $\av{p}^*{=} \int_{0}^1 dp~p P^*(p,t;p_0,0) =p_0^{-1} \int_0^1 dp~p^2 P(p,t;p_0,0)$ due to Eq.~(\ref{Bayes}). Then from Eq.~(\ref{FPEP}), we obtain 
 \be
 \frac{d\av{p^2}}{dt} = -\frac{1}{2 N} \int_0^1 dp p^2 \frac{\partial J(p,t)}{\partial p}
 \ee
 where, $J(p,t)=-\frac{1}{2 N}  \frac{\partial}{\partial p} \left( p(1-p) P \right)$ is the probability current. 
 On integrating the RHS of the above equation by parts and using the boundary conditions given in Eq.~(\ref{bcP}), we obtain an equation for the dynamics of  $\av{p}^*$ which can be easily solved, and 
 we find that  
 \be
\av{p(t)}^*{=} 1-(1-p_0)e^{- \frac{t}{N}} \label{cAM}
\ee 
The above expression shows that the mean frequency in the $A^*$ process grows {\it linearly} at short times ($t \ll N$); this is unlike when the $A$ allele is under positive selection and its conditional mean frequency rises exponentially \cite{Kaushik:2025}.  Note that the time scale $N_0$ does not appear in the dynamics of the arithmetic mean frequency.

The harmonic mean frequency conditioned on fixation is given by Eq.~(\ref{phmdef}) since 
\bea
\Bav{\frac{1}{p(t)}}^*=\int_{1/N}^1 dp \frac{1}{p} P^*(p,t;p_0,0) 
&=& \frac{1-\int_0^{1/N} dp P(p,t;p_0,0)}{p_0} =\frac{1-\Pi_0(t; p_0)}{p_0} \label{cHM}
\eea
on using that the distribution $P(p,t; p_0,0)$ is normalized to one, and due to Eq.~(\ref{pieces}). 
Since $\Pi_0(0; p_0)=0$ and $\Pi_0(\infty; p_0)=1-p_0$, it follows from Eq.~(\ref{cHM}) that the harmonic mean frequency eventually reaches one, starting from $p_0$. To understand the dynamics of the conditional mean of the inverse frequency, we use the results in Appendix~\ref{app_HM} and find that
\bsn
{\Bav{\frac{1}{p(t)}}^* \approx \label{aHM}}
p_0^{-1}  ~,&~$t \ll 2 N_0$ \label{aHM1} \\
\frac{1-(1-p_0) e^{-\frac{2 N_0}{t}}}{p_0} ~,&~$2 N_0 \ll t \ll 2 N$ \label{aHM2} \\
1+3 (1-p_0) e^{-\frac{t}{N}} ~,&~$t \gg 2 N$ \label{aHM3}
\esn
The above expression shows that the conditional harmonic mean frequency remains constant on a time scale proportional to the initial number of mutants in the $A^*$ process and approaches its asymptotic value on a time scale that grows linearly with the total population size $N$.

\begin{figure}[t]
     \centering
    \begin{subfigure}{0.49\textwidth}
         \centering
        \includegraphics[width=1.0\textwidth]{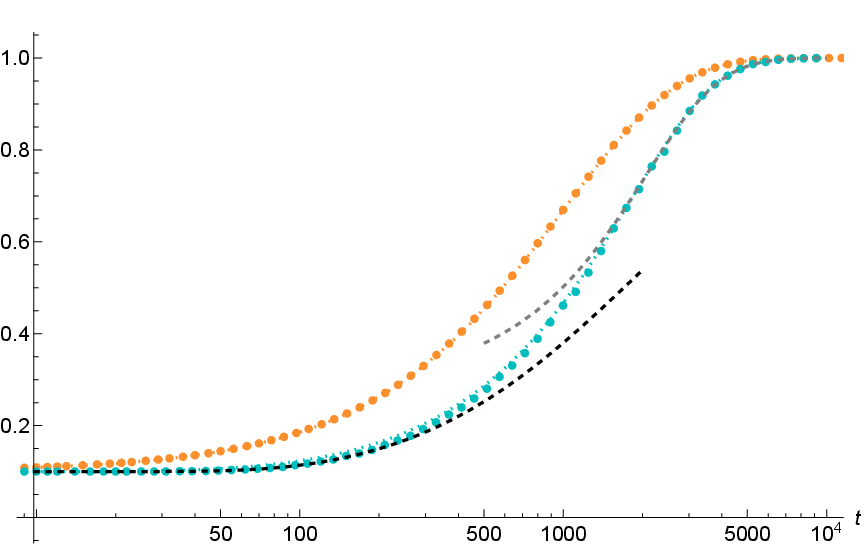}
         \caption{}
         \label{fig_AH}
     \end{subfigure}
     \begin{subfigure}{0.49\textwidth}
         \centering
         \includegraphics[width=1.0\textwidth]{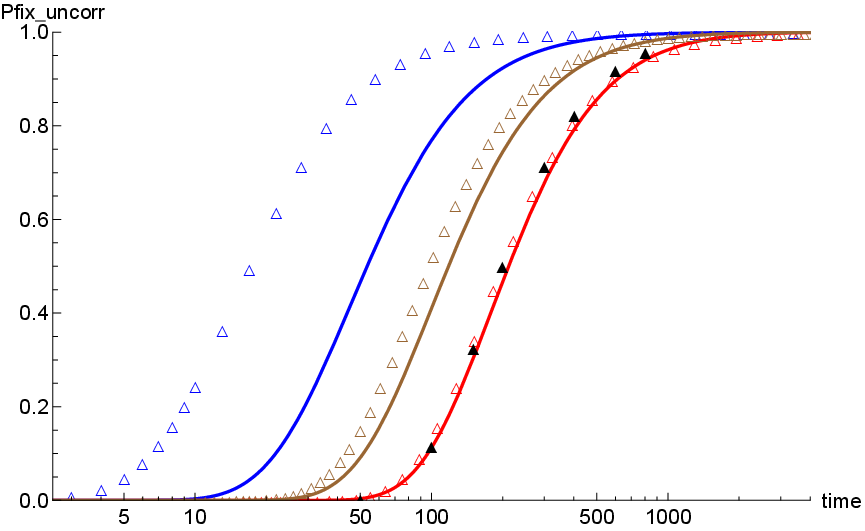}
         \caption{}
               \label{fig_pfixunco}
     \end{subfigure}
     \caption{(a) Arithmetic mean frequency (orange) and harmonic mean frequency (cyan) in the $A^*$ process obtained using numerical simulations (points) are compared with the exact expressions (solid lines) given by Eq.~(\ref{cAM}) and Eq.~(\ref{phmdef}), respectively, for $N=1000, p_0=0.1$. The RHS of Eq.~(\ref{phmdef}) is obtained by numerically integrating Eq.~(\ref{bFPE}) with appropriate boundary conditions. The approximate expressions for the conditional harmonic mean frequency obtained using   Eq.~(\ref{aHM2}) and Eq.~(\ref{aHM3}) are shown by black and gray dashed lines, respectively. (b) Scaled fixation probability, $\frac{\av{\Phi^{(u)}_1}^*}{x_0}$ of a single mutant allele $1$ in the uncorrelated model for $N=1000$, and $x_0^{-1}=N_0=10$ (blue), $50$ (brown) and $100$ (red) obtained from simulations ($\triangle$) and numerically solving Eq.~(\ref{uFPE}) (lines) is shown. The approximate result  ($\blacktriangle$) given by Eq.~(\ref{approx1})  is shown for $N_0=100$.}
       \label{fig_uncorr}
\end{figure}

 Figure~\ref{fig_AH} shows that Eq.~(\ref{phmdef}) and Eq.~(\ref{cAM}) agree with the corresponding results obtained using numerical simulations,  and as expected, the conditional mean frequency is larger than the corresponding harmonic mean frequency. The harmonic mean frequency obtained using the approximations in Eq.~(\ref{aHM}) is also compared with the simulation results, and we find a reasonable agreement in their respective regimes of validity. 

 \subsection{Fixation probability}
 
 From Eq.~(\ref{Pi1F}) and Eq.~(\ref{uncorr1}), we obtain the fixation probability averaged over the uncorrelated population size to be
\bea
\av{\Phi^{(u)}_1(t; x_0)}^*=x_0+x_0(1-x_0)  \sum_{n=1}^\infty   \frac{(-1)^n (2 n+1)}{n} P^{(1,1)}_{n-1}(1-2 x_0) e^{-\Lambda_n \int_0^t dt_1 \bav{\frac{1}{p(t_1)}}^*} \label{Pi1Fun}
\eea
Thus, similar to Eq.~(\ref{bFPE}) for constant population size,  the above fixation probability is a solution of the following backward FPE, 
\bea
\frac{\partial \av{\Phi^{(u)}_1}^*}{\partial t} &=&  \Bav{\frac{1}{p(t)}}^* \frac{x_0 (1-x_0)}{2 N} \frac{\partial^2 \av{\Phi^{(u)}_1}^*}{\partial x_0^2} \label{uFPE}
\eea
with boundary conditions, $\av{\Phi^{(u)}_1(t; 0)}^*=0, \av{\Phi^{(u)}_1(t; 1)}^*=1$, and initial condition $\av{\Phi^{(u)}_1(0; x_0)}^*=0$; the results obtained by numerically integrating the above equation are shown in Fig.~\ref{fig_pfixunco}. 
We also measured the probability $\av{\Phi^{(u)}_1}^*$ in simulations in which 
  a new stochastic trajectory of the subpopulation $A$ that eventually fixes was generated at {\it each} generation of the Wright-Fisher process for the mutant allele $1$ so that the subpopulation size at all times are uncorrelated. As shown in Fig.~\ref{fig_pfixunco}, the simulation results for small $N_0$ are not captured by Eq.~(\ref{uFPE}) for reasons already discussed in Sec.~\ref{FPEcst}.

But for sufficiently large $N_0$, on using Eq.~(\ref{aHM2}) for $t \ll 2 N$ in Eq.~(\ref{Pi1Fun}), we obtain
\be
\av{\Phi^{(u)}_1}^* \approx x_0+x_0(1-x_0)  \sum_{n=1}^\infty   \frac{(-1)^n (2 n+1)}{n} P^{(1,1)}_{n-1}(1-2 x_0) e^{-\frac{n (n+1) t}{2 N_0} [1-\left(1-p_0\right) e^{-\frac{2 N_0}{t}}+\frac{2 N_0}{t} \left(1-p_0\right) \Gamma
   \left(0,\frac{2 N_0}{t}\right)]}  \label{approx1}
\ee
which is in good agreement with the simulation results shown in Fig.~\ref{fig_pfixunco}.
At longer times $t \gg 2 N$, the dynamics are determined by the smallest eigenvalue $\Lambda_1$, and as shown in Appendix~\ref{app_approach}, the fixation probability approaches its asymptotic value as 
\bea
x_0-\av{\Phi^{(u)}_1}^* &\approx& 3 x_0(1-x_0) e^{-\frac{1}{N} \int_0^t dt_1 \av{\frac{1}{p(t_1)}}^*} \\
&\approx& 3 x_0(1-x_0) e^{-3 (1-p_0)} e^{-\frac{t}{N}}
\eea
 As expected and in agreement with the simulation results in Fig.~\ref{fig_appro}, the approach to the eventual fixation probability occurs over a time scale $N$. The above equations also show that the amplitude of the exponential decay depends on both $p_0$ and $x_0$, and the simulation data shown in Fig.~\ref{fig_appro} for two values of $x_0$ is consistent with the above prediction.

\section{Correlations in the $A^*$ process}
\label{CorrM}

As Fig.~\ref{fig_fullun} and Fig.~\ref{fig_appro} show, the uncorrelated population size model does not accurately capture the fixation probability in the full model, and is a particularly poor approximation at intermediate times. 
Thus the temporal correlations in Eq.~(\ref{cumexpn}) can not be ignored; below we study these correlations numerically and analytically using FPEs for sufficiently large $N_0$, and also discuss how they affect the fixation probability.

 \subsection{Two time correlation function}

We first consider the unconnected two-time correlation function for the inverse population frequency in the $A^*$ process. For $t_2 \geq t_1$, we can write
\bea
\Bav{\frac{1}{p(t_2) p(t_1)}}^* &=& \int_{1/N}^1 \frac{dp}{p} \int_{1/N}^1 \frac{dp'}{p'} P^*(p',t_2|p,t_1) P^*(p,t_1|p_0,0) \\
&=& \int_{1/N}^1  \frac{dp}{p} P^*(p, t_1|p_0,0) \times \frac{1-\Pi_0(t_2-t_1; p)}{p} \\
&=& \frac{1}{p_0} \int_{1/N}^1 \frac{dp}{p} P(p,t_1|p_0,0) [1-\Pi_0(t_2-t_1; p)] \label{HMcor}
\eea
on using Eq.~(\ref{Bayes}) and Eq.~(\ref{cHM}), and where $P(p,t_1|p_0,0)$ is given by Eq.~(\ref{pieces}). Due to the nonzero lower limit of integration, the first term on the RHS of Eq.~(\ref{pieces}) does not contribute to the above integral. 
Then due to  the second term on the RHS of Eq.~(\ref{pieces}) and the boundary condition $\Pi_0(t; 1)=0$, we obtain
\bea
\Bav{\frac{1}{p(t_2) p(t_1)}}^* &=& (1-p_0) \int_{1/N}^1 dp  \frac{f(p, t_1|p_0,0)}{p_0 (1-p_0)} \frac{1-\Pi_0(t_2-t_1; p)}{p}+ \frac{\Pi_1(t_1;p_0)}{p_0}  \label{HMcor2} 
\eea
Using the above equation and Eq.~(\ref{cHM}), the connected correlation function
 \be
 C(t_2, t_1)=\Bav{\frac{1}{p(t_2) p(t_1)}}^*-\Bav{\frac{1}{p(t_2)}}^* \Bav{\frac{1}{p(t_1)}}^*
 \ee 
 can be obtained.

 \subsubsection{Variance of the inverse frequency}
 
For $t_1=t_2=t$, due to the initial condition $\Pi_0(0; p_0)=0$, Eq.~(\ref{HMcor2}) reduces to
 \bea
\Bav{\frac{1}{p^2(t)}}^* &=& (1-p_0) \int_{1/N}^1 dp  \frac{f(p, t|p_0,0)}{p_0 (1-p_0) p} + \frac{\Pi_1(t;p_0)}{p_0}  \label{vardef}
\eea
For $t \ll 2 N$, as explained in Appendix~\ref{app_HM}, the fixation probability $\Pi_1 \approx 0$  so that the last term on the RHS of the above equation can be ignored. Furthermore, as shown in Appendix~\ref{app_var}, for $p_0 \to 0, t \ll 2 N$, 
\bea
\Bav{\frac{1}{p^2(t)}}^* &\approx& (1-p_0) \int_{1/N}^1 dp \frac{f(p, t|p_0,0)}{p_0 (1-p_0)} \frac{1}{p} \\
&\approx&
\frac{y}{p_0^2} {\left[e^{-y} \left(y \text{Ei}(y)-y \ln \left(\frac{{2 y}}{t}\right)+1 \right)-1 \right]} \label{varA1}
\eea
where $y=\frac{2 N_0}{t}$ and $\text{Ei}(x)= -P [\int_{-x}^\infty dz \frac{e^{-z}}{z}]$ is the exponential integral [see (6.2.6) of \cite{DLMF}]. 
For $t \ll 2 N_0$, using (6.12.2) of \cite{DLMF} for the asymptotic expansion of $\text{Ei}(x)$, we find that the RHS of Eq.~(\ref{varA1}) can be approximated by $p_0^{-2} (1+\frac{t}{N_0})$. But for $2 N_0 \ll t \ll 2 N$, using (6.6.1) of \cite{DLMF} for the power series expansion  of $\text{Ei}(x)$,  we obtain $\bav{\frac{1}{p^2(t)}}^* \approx \frac{y^2}{p_0^2} \ln \left( \frac{t}{2} \right)=\left(\frac{2 N}{t} \right)^2 \ln  \left( \frac{t}{2} \right)$. Thus, at intermediate times, the second moment of the inverse frequency in the $A^*$ process decays algebraically and is independent of $N_0$. At larger times where $t \gg 2 N$, using the results in Appendix~\ref{app_HM} and Appendix~\ref{app_var}, we obtain $\bav{\frac{1}{p^2(t)}}^* \approx 1+6 e^{-\frac{t}{N}} \ln N$. 

\begin{figure}[t]
     \centering
    \begin{subfigure}{0.49\textwidth}
         \centering
         \includegraphics[width=1.0\textwidth]{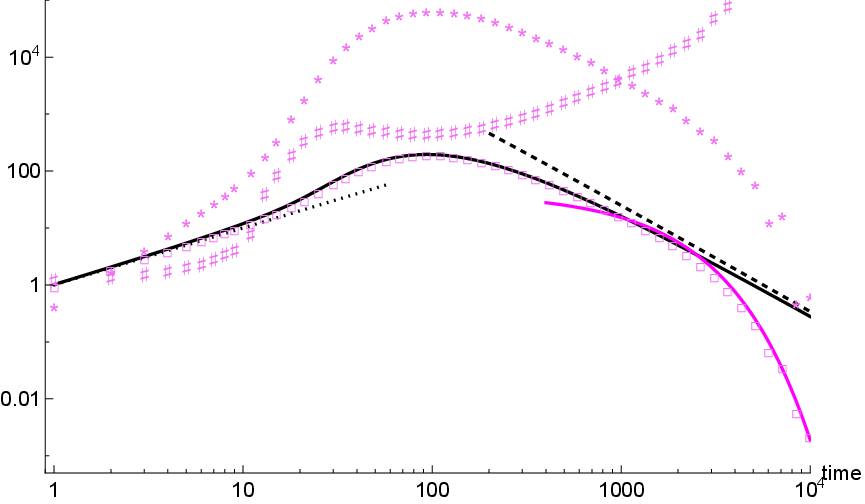}
         \caption{}
         \label{fig_mom}
     \end{subfigure}
     \begin{subfigure}{0.49\textwidth}
         \centering
         \includegraphics[width=1.0\textwidth]{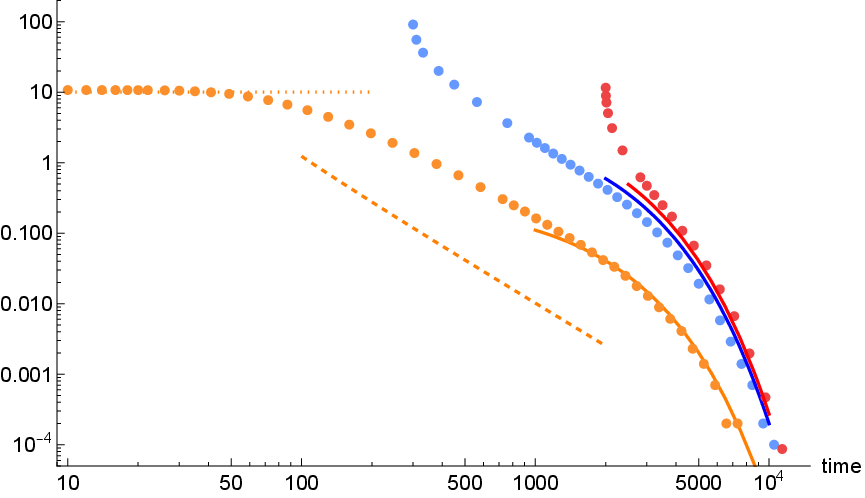}
         \caption{}
               \label{fig_corr2}
     \end{subfigure}
     \caption{(a) Variance ($\Box$), skewness ($\star$) and scaled kurtosis ($\sharp$) of the inverse frequency in the $A^*$ process as a function of time for $N=1000, N_0=100, x_0=N_0^{-1}$ obtained from simulations is shown. The solid lines  show the variance obtained using Eq.~(\ref{varA1}) and Eq.~(\ref{aHM2}) for $t \ll 2 N$ and Eq.~(\ref{k23}) for $t \gg 2 N$;  the approximate expressions Eq.~(\ref{k21}) and Eq.~(\ref{k22}) are depicted by dotted and dashed line, respectively. (b) Unequal time correlation function, $C(t_2, t_1)=\bav{\frac{1}{p(t_2) p(t_1)}}_c^*$  as a function of time $t_2$ for  $t_1=10$ (orange), $300$ (blue), $2000$ (red) is obtained from simulations (points). The dotted line shows Eq.~(\ref{k21}) evaluated at time $t_1=10$ and the dashed line shows $0.01 (t_2-t_1)^{-2}$ to test the scaling in Eq.~(\ref{corrInt2}) for $t_1=10$;  the solid lines show Eq.~(\ref{cexpo}) in their respective regions of validity.  }
       \label{fig_corr24}
\end{figure}

 Using the above approximations for the second moment and Eq.~(\ref{aHM}) for the mean of the inverse frequency, we find that the dynamical behavior of the variance falls in three distinct regimes: initial linear increase, algebraic decay as $t^{-2}$ and finally an exponential approach to zero:
 \bsn
 {\kappa_2(t)=\Bav{\frac{1}{p^2(t)}}^*-{\Bav{\frac{1}{p(t)}}}^{*2} \approx}
 \frac{t}{N_0 p_0^2} ~,&~ $t \ll 2 N_0$ \label{k21} \\
 \left(\frac{2 N}{t} \right)^2 \ln  \left( \frac{t}{2} \right) ~,&~ $2 N_0 \ll t \ll 2 N$ \label{k22}\\
 6 e^{-\frac{t}{N}} \ln N ~,&~$t \gg 2 N$ \label{k23}
 \esn
 The above expression also shows that the maximum in the variance occurs at time $\sim N_0$. Figure~\ref{fig_mom} shows a comparison between the simulation results and the above approximations for the variance. We find that for $t \ll 2 N$, the variance obtained using Eq.~(\ref{varA1}) and Eq.~(\ref{aHM2}) is in good agreement with the simulations; the initial linear growth indicated by Eq.~(\ref{k21}) is also observed but the approximate expression given by Eq.~(\ref{k22}) at intermediate times is seen for a very short time span as the exponential decay given by Eq.~(\ref{k23}) sets in. 
 
 \subsubsection{Unequal time correlation function}

For $t_2 > t_1$, we first note that, as discussed in Sec.~\ref{FPEcst}, $\Pi_0(t_2-t_1; p) \approx 0$ for $t_2-t_1 \ll 2 N p$. Then for a given $t_1 \ll 2 N_0$, the distribution $f(p, t_1|p_0, 0)$ is expected to remain close to the initial frequency $p_0$, and therefore, for $t_1 \ll 2 N_0, t_2 -t_1 \ll 2 N_0$, the unequal time correlation function given by Eq.~(\ref{HMcor2}) can be approximated by the second moment at time $t_1$ [see Eq.~(\ref{vardef})]. 
But at larger times ($t_2 - t_1 \gg 2N_0$) where the extinction probability $\Pi_0$ is non-negligible, the correlation function depends on time $t_2$ also. The simulation data in Fig.~\ref{fig_corr2} for $t_1=10$ shows that indeed $C(t_2, t_1) \approx \kappa_2(t_1)$ when $t_2 \ll 2 N_0$ and decreases thereafter.  

Then, as shown in Appendix~\ref{app_two1}, except when both $t_2, t_1 \lesssim 2 N_0$, the unconnected two-time correlation function defined in Eq.~(\ref{HMcor2}) can be rewritten as 
\bea 
&&\Bav{\frac{1}{p(t_2) p(t_1)}}^*= \Bav{\frac{1}{p(t_1)}}^* \nn \\
&+& (1-p_0) \sum_{n, m=1}^\infty \frac{(2 n+1) (n+1)}{n} P^{(1,1)}_{n-1}(1-2 p_0) e^{-\Lambda_{n} t_1} \frac{2 m+1}{m} 
e^{-\Lambda_{m} (t_2-t_1)} \frac{\min\{m,n\}}{\max\{m,n\}+1} \label{unequalsum}
\eea
As for variance, due to the exponential factors in the above summand, one can consider cases when $t_1$ and $t_2-t_1$ are smaller or larger than $2 N$ to obtain approximate expressions for the unequal time correlation function. 

We first consider the parameter regime where $t_1 \ll 2 N_0$ but $2 N_0 \ll t_2 -t_1 \ll 2 N$. Then, as discussed in Appendix~\ref{app_two22}, we obtain
\be
\Bav{\frac{1}{p(t_2) p(t_1)}}^*  \propto \frac{N^2}{(t_2-t_1)^2} \label{corrInt2}
\ee
which is in reasonable agreement with the simulation results shown in Fig.~\ref{fig_corr2} for $t_1=10$; a better quantitative agreement seems difficult to obtain due to the crossovers on either side of the intermediate regime, as mentioned above for the variance.  For $t_1 \ll 2 N_0$ and $t_2-t_1 \gg 2 N$, see Eq.~(\ref{cexpo1}) below. 

For $t_1 \gg 2 N_0$, as Fig.~\ref{fig_corr2} shows, much of the dynamics of the correlation function can be captured by its  late time behavior where $t_2 -t_1 \gg 2 N$. Then as described in Appendix~\ref{app_two2}, we obtain
\bsn
{C(t_2, t_1) \approx \label{cexpo}}
\frac{3 t_1}{N_0} e^{-\frac{t_2}{N}} ~,&~ $t_1 \ll 2 N_0$ \label{cexpo1} \\
\frac{3 t_1}{N_0} (1-e^{-\frac{2 N_0}{t_1}}) e^{-\frac{t_2}{N}} ~,&~ $2 N_0 \ll t_1 \ll 2 N$\\
6 e^{-\frac{t_2}{N}}~,&~ $t_1 \gg 2 N$
\esn
These expressions  are found to be  in good agreement with the simulation results shown in Fig.~\ref{fig_corr2} for various $t_1$.

 \subsection{Skewness and kurtosis}
 
 As described in the last subsection, the correlation function depends not only on the time difference but also on the earlier time $t_1$, and therefore the  $A^*$ process for the inverse frequency is not a stationary process. We now ask if this process is a Gaussian process; here, we do not address this question analytically and instead numerically measure the skewness and scaled kurtosis defined as 
 \bea
 \kappa_3(t) &=& \av{r^3(t)}^*\\
\kappa_4(t) &=& \frac{\av{r^4(t)}^*-3 \av{r^2(t)}^{*2}}{3 \kappa^2_2(t)}
\eea
where, $r(t)= \frac{1}{p(t)}-\bav{\frac{1}{p(t)}}^*$. 
For a stationary Gaussian process, $\kappa_3(t)=0, \kappa_4(t)=1$. But as the simulation results in Fig.~\ref{fig_mom} show, $\kappa_3(t)$ is nonzero, and, as for variance, it is also a nonmonotonic function of time with an algebraic decay at intermediate times. The scaled kurtosis is seen to be close to one at short times but it is also a nonmonotonic function of time. As the kurtosis must vanish at large times when the allele $A$ has fixed, an increase in $\kappa_4$ for $t \gtrsim 1000$ is likely a numerical artefact. 
We thus conclude that the $A^*$ process for the inverse frequency is not a Gaussian process.

 \subsection{Effect of correlations on the fixation probability}
 \label{sec_fullfix}

\begin{figure}[t]
     \centering
    \begin{subfigure}{0.49\textwidth}
         \centering
         \includegraphics[width=1.0\textwidth]{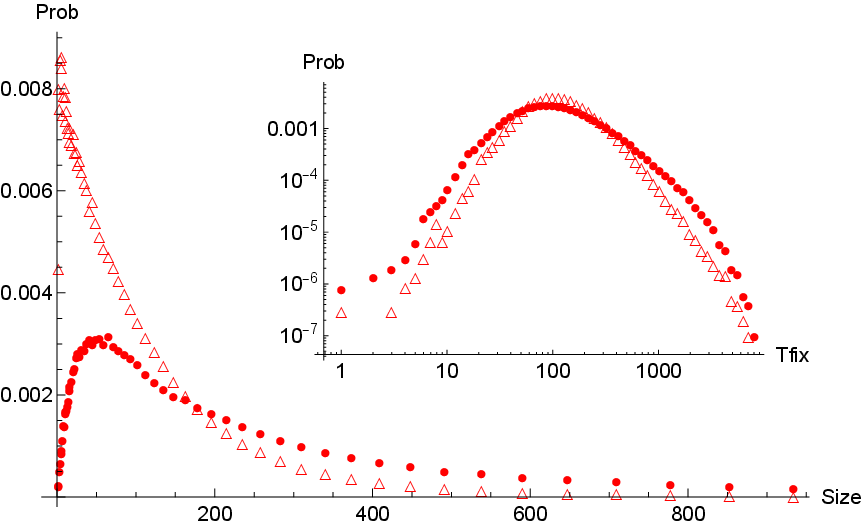}
         \caption{}
         \label{fig_Nf}
     \end{subfigure}
     \begin{subfigure}{0.49\textwidth}
         \centering
         \includegraphics[width=1.0\textwidth]{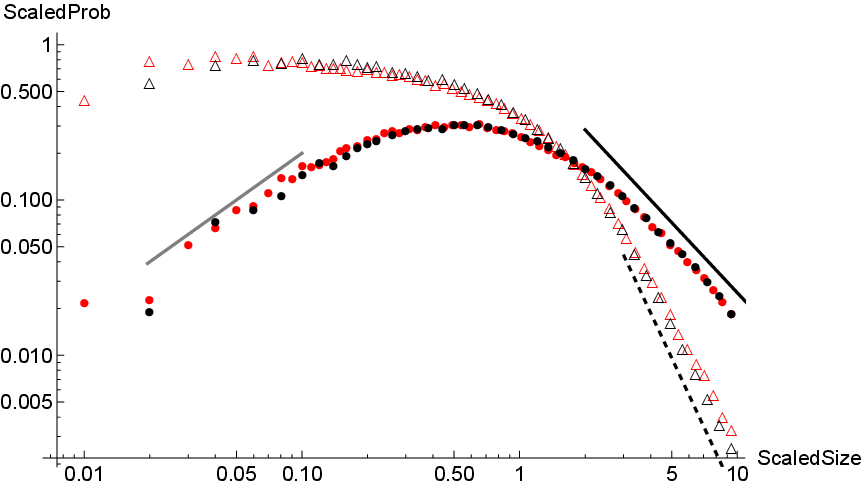}
         \caption{}
               \label{fig_Nf2}
     \end{subfigure}
     \caption{(a) Distribution of population size $N_f$ at the time of fixation of allele $1$ (main) and distribution of fixation time (inset)  for $N=1000, N_0=100$ obtained from simulation for the full model ($\bullet$) and uncorrelated model ($\triangle$). (b) Scaled distribution, $N_0 Q(N_f)$ as a function of scaled size, $\frac{N_f}{N_0}$ for $N=1000, N_0=100$ (red) and $N=500, N_0=50$ (black) for the full model ($\bullet$) and uncorrelated model ($\triangle$). The slope of the solid lines are  $1$ (gray) and $-3/2$ (black), and that of dashed line is $-3$.}
       \label{fig_size}
\end{figure}

We first note that as displayed in Fig.~\ref{fig_fullun} for $N_0=100$, the fixation probability in the full model differs from that in the uncorrelated model  for $100 \lesssim t \lesssim 1000$ which, as Fig.~\ref{fig_corr24} shows, is also the time range where variance and unequal time correlation function $C(t_2, t_1)$ for $t_1, t_2 \ll 1000$ decay  algebraically.  Thus, in general, we expect that the fixation probability in the full model can not be approximated by that in the  uncorrelated model at intermediate times where $2N_0 \ll t \ll 2N$. 

Furthermore, Fig.~\ref{fig_fullun} also shows that in the full model, the fixation probability on these time scales is {\it smaller} than that in the uncorrelated model suggesting that the population size is effectively larger in the former case. To understand this, we measured the probability $Q(N_f)$ that the population size in the $A^*$ process is $N_f$ at the time when allele $1$ fixes. As Fig.~\ref{fig_Nf} shows, this distribution is heavily skewed towards small $N_f$ in the uncorrelated model; 
however, this does not mean that when correlations are neglected, the fixation occurs at very small times as the distribution of fixation time in the uncorrelated model and full model are quite similar, see the inset of Fig.~\ref{fig_Nf}. 
Rather when correlations are absent, the population size fluctuates substantially between generations and therefore, even at large times, its size can be small (although the mean size increases according to Eq.~(\ref{cAM})). 
In contrast, due to correlations  in the 
full model, the population size increases in a smoother fashion and allele  $1$ typically encounters a large population 
(see Fig.~\ref{fig_Nf}).  

Figure~\ref{fig_Nf2} suggests that for large $N_f$, the distribution $Q(N_f)$ decreases algebraically, as $N_f^{-3}$ in the uncorrelated model and $N_f^{-3/2}$ in the full model. The behavior in the former case can be obtained analytically as explained in Appendix~\ref{app_Nf}, and we find that the population size distribution is of the following scaling form, 
\bea
Q^{(u)}(N_f)=\frac{1}{N_0} {\cal Q}^{(u)} \left( \frac{N_f}{N_0}\right)
\eea
where, the scaling function, ${\cal Q}^{(u)}(x)$ is a constant for $x \ll 1$ and decays as $x^{-3}, x \gg 1$. 
As the data collapse in Fig.~\ref{fig_Nf2} indicates, the above scaling form holds for the full model also but the scaling function changes nonmonotonically; a numerical fit suggests that the scaling function initially increases as $x$ and then decays as $x^{-3/2}$.

\section{Discussion}
\label{sec_disc}

Natural populations have a finite carrying capacity, but their size does not remain fixed at the maximum possible size and, in general,  it changes stochastically in a correlated fashion. However, most work in population-genetic studies of biological evolution implicitly assume that the population sizes are uncorrelated, and subsume the effect of changing size in an effective population size \cite{Waples:2022} given by the harmonic mean of the varying population size (see Eq.~(\ref{phmdef})); however, an effective population size does not exist when the correlations can not be neglected \cite{Sjodin:2005,Eriksson:2010,Jain:2025}. In a similar vein, here we have shown that  the chance of fixation is strongly affected during the time regime when correlations in the population size are substantial.

Unfortunately, we are unable to obtain exact expression for the time-dependent fixation probability as the $A^*$ process which is the conditional Wright-Fisher process of our interest is found to be neither a stationary nor a Gaussian process; an example where exact results for the dynamics of fixation probability have been obtained is given in \cite{Jain:2025} where  the population size  follows a random telegraph process, which is a stationary process and where the correlations in the inverse population size decay exponentially. However, such analytically tractable models make adhoc assumptions for the variation in the population size \cite{Maruyama:1985c,Parsons:2010,Waxman:2012,Asker:2025,Jain:2025}, while 
the stochastic process governing the fluctuations in the population size considered here arises naturally from a well established model of genetic hitchhiking \cite{Smith:1974}.

Here, we have discussed the dynamics of fixation probability when the $A$ subpopulation is neutral. But  when the subpopulation is under selection (and therefore, grows exponentially), although the eventual fixation probability has been studied \cite{Parsons:2007a,Engen:2009,Czuppon:2018},  to our knowledge, the dynamics have not been investigated 
and should be addressed. In this work, the analytical results are obtained when the initial size of the $A$ subpopulation is of the order of the total population size because the  Fokker-Planck equations analyzed here do not hold otherwise (see Fig.~\ref{fig_extN} and Fig.~\ref{fig_pfixunco}). However, in biologically relevant situations, the initial size $N_0 \sim {\cal O}(1)$ and  one needs to work with the discrete number of individuals (and not continuous frequency) at 
 short times; a detailed analyses, perhaps along the lines of  \cite{Parsons:2010}, is desirable. \\

\noindent{\it Acknowledgments:} HS thanks JNCASR for support through the SRFP-2024 and DST for funding through the INSPIRE-SHE, and is grateful to Venu Goswami, Ayantik Kundu, Prashant Singh and 
Lakshita Jindal for helpful discussions. 

\clearpage
\appendix
\makeatletter
\renewcommand{\@seccntformat}[1]{Appendix \csname the#1\endcsname\quad}
\makeatother
\renewcommand{\thesection}{\Alph{section}}
\numberwithin{equation}{section}

\section{General model}
\label{app_genl}

To understand how other sites linked to selectively neutral regions of the genome affect the neutral genetic diversity \cite{Charlesworth:2021}, one considers a finite population of binary sequences denoted by $\{\sigma_1, ..., \sigma_L; \eta_1, ..., \eta_\ell\}$  where $\sigma_i=a, A$ and $\eta_i=0, 1$, and $L, \ell \gg 1$. It is usually assumed that (i) the sites labeled by $\sigma_i$ are, in general, under selection and the $\eta_i$-sites are neutral, (ii) for long sequences and low mutation rates, at most one mutation occurs at a site from wildtype allele ($a$ or $0$) to mutant allele ($A$ or $1$) and the reverse mutations can be neglected  \cite{Kimura:1969}, and (iii) the linkage between the sites can be broken due to genetic recombination; for a recent work on models that incorporate such and other biological details, see \cite{Kaushik:2025} and references therein. Here, as we are mainly interested in understanding the effect of fluctuating population size on the fixation probability of a mutant at a fully linked neutral site, we have focused on a model with $L=\ell=1$ and assumed that all the four sequence configurations are equally fit. 

\section{Absorption probabilities in constant-sized population}
\label{app_HM}


Here, we analyze the extinction and fixation probability in a population of fixed size $N$. First, consider the sum on the RHS of Eq.~(\ref{Pi0}) for extinction probability,
\bea
\Sigma_0=\sum_{n=1}^\infty   \frac{2 n+1}{n} P^{(1,1)}_{n-1}(1-2 p_0) e^{- \frac{n(n+1) t}{2 N}}
\eea
on using Eq.~(\ref{eval}). For $t \gg 2 N$, it is sufficient to consider the term corresponding to the lowest eigenvalue ($n=1$) which yields 
\be
\Sigma_0 \stackrel{t\gg 2 N}{\approx} 3 e^{-\frac{t}{N}} \label{HMlate}
\ee
since the Jacobi polynomial $P^{(\alpha,\beta)}_0(y)=1$. 
On the other hand, for $t \ll 2 N$, $n \gg 1$ also contribute to the sum $\Sigma_0$. Then it is useful to approximate the Jacobi polynomials as \cite{Kaushik:2025}
\bea
P_{m-1}^{(1,1)}(1-2x) &{=}& m \sum_{n=0}^{m-1} \frac{(1-m)_n (2+m)_n}{(2)_n}\frac{x^n}{n!}\\ &\stackrel{m \gg 1, x \to 0}{\approx}& \sum_{n=0}^\infty \frac{(-1)^n (m^{2} x)^n}{n! \Gamma (n+2)} \\
&=& \frac{J_1\left(2 m \sqrt{x}\right)}{\sqrt{x}} \label{PJapprox}
\eea
which is obtained in the scaling limit $m \to \infty, x \to 0, m \sqrt{x}$ finite, and where $J_1(y)$ is the Bessel function of the first kind. For small $p_0$, we can then write
\bea
\Sigma_0 &{\approx}& 2 \int_0^\infty dn  \frac{J_1(2 n \sqrt{p_0})}{\sqrt{p_0}} e^{-\frac{n^2}{2 N} t} \\
&=& \frac{1-e^{-\frac{2 N_0}{t}}}{p_0} ~,~ t \ll 2N \label{HMearl}
\eea
which is valid when $t, N_0 \gg 1$ and $\frac{t}{2 N_0}$ are finite. Thus, as discussed in Sec.~\ref{FPEcst}, the results obtained from the FPE agree with those from simulations if the initial number of mutants and time are  sufficiently large. Equation~(\ref{HMearl}) also shows that for $t \ll 2 N_0$, as $\Sigma_0 \approx p_0^{-1}$, the extinction probability $\Pi_0$ in Eq.~(\ref{Pi0}) is negligible; furthermore, using Eq.~(\ref{HMearl}) in Eq.~(\ref{cHM}), we obtain Eq.~(\ref{aHM1}) and Eq.~(\ref{aHM2}), and similarly, Eq.~(\ref{HMlate}) leads to Eq.~(\ref{aHM3}).

Next, consider the sum on the RHS of Eq.~(\ref{Pi1}) for fixation probability,
\be
\Sigma_1=\sum_{n=1}^\infty   \frac{(-1)^n (2 n+1)}{n} P^{(1,1)}_{n-1}(1-2 p_0) e^{- \frac{n(n+1) t}{2 N}} 
\ee
As for $\Sigma_0$, for $t \gg 2 N$, it is sufficient to keep the term corresponding to the lowest eigenvalue in the above sum, and we have $\Sigma_1 \approx -(1-p_0)^{-1} e^{-\frac{t}{N}}$. But even at shorter times where $t \ll 2 N$, our numerical study of the above sum shows that the alternating series $\Sigma_1$ converges rapidly and can be approximated by first few terms in the summand of $\Sigma_1$. We also find that the sum remains close to its initial value $-(1-p_0)^{-1}$ until $t \lesssim N$, in accordance with the conclusion  in the main text  that the initial number $N_0$ does not play a role in the dynamics of $\Pi_1$. 

\section{Long time dynamics in the uncorrelated model}
\label{app_approach}

For the uncorrelated model, on retaining only the term corresponding to the lowest eigenvalue in  Eq.~(\ref{Pi1Fun}), we get 
\bea
\av{\Phi^{(u)}_1}^* \approx x_0-3 x_0(1-x_0) e^{-\frac{1}{N} \int_0^t dt_1 \bav{\frac{1}{p(t_1)}}^*} 
\eea
since $P^{(1,1)}_{0}(x)=1$. Then, due to Eq.~(\ref{Pi0}), we obtain
\bea
\frac{1}{N}\int_0^t dt_1 \Bav{\frac{1}{p(t_1)}}^* &=& \frac{t}{N} +\left(1-p_0\right) \sum _{n=1}^{\infty } \frac{2 (2 n+1) P_{n-1}^{(1,1)}\left(1-2 p_0\right) \left(1-e^{-\frac{n (n+1) t}{2 N}}\right)}{n^2 (n+1)} \\
&\approx& \frac{t}{N} +\left(1-p_0\right) \sum _{n=1}^{\infty } \frac{2 (2 n+1) P_{n-1}^{(1,1)}\left(1-2 p_0\right) }{n^2 (n+1)}
\eea
so that
\bea
x_0-\av{\Phi^{(u)}_1}^* &\approx& 3 x_0(1-x_0) e^{-\left(1-p_0\right) \sum _{n=1}^{\infty } \frac{2 (2 n+1) P_{n-1}^{(1,1)}\left(1-2 p_0\right) }{n^2 (n+1)}} e^{-\frac{t}{N}} \\
&\approx& 3 x_0 (1-x_0) e^{-3 (1-p_0)} e^{-\frac{t}{N}}
\eea

\section{Approximations for the variance of the inverse frequency}
\label{app_var}

Consider the integral on the RHS of Eq.~(\ref{vardef}) which can be written as 
\bea
\int_{1/N}^1 dp \frac{f(p, |p_0,0)}{p_0 (1-p_0)} \frac{1}{p} 
&=& \sum_{n=1}^\infty   \frac{(2 n+1) (n+1)}{n} P^{(1,1)}_{n-1}(1-2 p_0) e^{-\Lambda_{n} t}\int_{1/N}^{1-\frac{1}{N}} dp \frac{P^{(1,1)}_{n-1}(1-2 p)}{p} \label{app_var1}
\eea
on using Eq.~(\ref{fpt}). We first note that for $t \ll 2 N$, most of the contribution to the integral on the RHS comes from 
$p \to 0$ and to the sum from $n \gg 1$. 
Then using the approximation in Eq.~(\ref{PJapprox}) for the Jacobi polynomial in the above equation and on approximating the sum by an integral, we obtain
\bea
\int_{1/N}^1 dp \frac{f(p, t|p_0,0)}{p_0 (1-p_0)} \frac{1}{p} &\approx& \int_0^\infty dn \; 2 n \; \frac{J_1\left(2 n \sqrt{p_0}\right)}{\sqrt{p_0}} e^{-\frac{n^2 t}{2 N}}  \times 4 n  \int_{\frac{2 n}{\sqrt{N}}}^{2 n} dy \frac{J_1(y)}{y^{2}} \\
&\approx& \frac{1}{p_0^2} \int_0^\infty dz z^2 J_1(z) e^{-\frac{z^2 t}{8 N p_0}} \int_{\frac{z}{\sqrt{N p_0}} \to 0}^{\frac{z}{\sqrt{p_0}} \to \infty} dy \frac{J_1(y)}{y^{2}} \label{appDint2}\\
&\approx& \frac{1}{4 p_0^2} \int_0^\infty dz z^2 J_1(z) e^{-\frac{z^2 t}{8 N p_0}} \ln \left(\frac{{4 N p_0}}{z^2} \right) \label{appDint}
\eea
where the last expression is obtained on using that the lower limit in the inner integal in Eq.~(\ref{appDint2}) scales as $t^{-1/2} \stackrel{t \gg 1}{\to} 0$ while the upper limit $\sim \sqrt{\frac{N}{t}} \gg 1$ as we are working in the regime where $t \ll 2 N$. On performing the integral in Eq.~(\ref{appDint}), we finally obtain Eq.~(\ref{varA1}) in the main text. For $t \gg 2 N$, it is sufficient to evaluate the sum on the RHS of (\ref{app_var1}) with $n=1$ which yields $6 e^{-\frac{t}{N}} \ln N$. 

\section{Unequal time correlation function}
\label{app_two1}

Consider the integral on the RHS of Eq.~(\ref{HMcor2}) when either $t_1 \ll 2 N_0, t_2 \gg 2 N_0$ or $t_2 > t_1 \gg 2 N_0$ so that the extinction probability $\Pi_0(t_2-t_1; p)$ is not negligible. Then due to Eq.~(\ref{Pi0}) and  Eq.~(\ref{fpt}), the integrand 
\bea
 &&\frac{f(p,t_1|p_0,0)}{p_0 (1-p_0)} \frac{1-\Pi_0(t_2-t_1; p)}{p} \\
 &=&  \sum_{n=1}^\infty   \frac{(2 n+1) (n+1)}{n} P^{(1,1)}_{n-1}(1-2 p_0) P^{(1,1)}_{n-1}(1-2 p) e^{-\Lambda_{n} t_1} [1+(1-p)  \sum_{m=1}^\infty   \frac{2 m+1}{m} P^{(1,1)}_{m-1}(1-2 p) e^{-\Lambda_{m} (t_2-t_1)}] \nn
\eea
On integrating both sides of the above equation over $p$, we get
\bea
&& \int_{1/N}^1 dp \frac{f(p,t_1|p_0,0)}{p_0 (1-p_0)} \frac{1-\Pi_0(t_2-t_1; p)}{p} \\
&\approx& \int_{0}^1 dp \frac{f(p,t_1|p_0,0)}{p_0 (1-p_0)} \frac{1-\Pi_0(t_2-t_1; p)}{p} \\
&=& \sum_{n=1}^\infty   \frac{2 n+1}{n} P^{(1,1)}_{n-1}(1-2 p_0) (1-(-1)^n) e^{-\Lambda_{n} t_1} \nn \\
&+& \sum_{n,m=1}^\infty \frac{(2 n+1) (n+1)}{n} P^{(1,1)}_{n-1}(1-2 p_0) e^{-\Lambda_{n} t_1} \frac{2 m+1}{m} 
e^{-\Lambda_{m} (t_2-t_1)} \frac{\min\{m,n\}}{\max\{m,n\}+1} \label{integral1}
\eea
where we have used that 
\bea
\int_0^1 dx  P^{(1,1)}_{n-1}(1-2x) &=& \frac{1-(-1)^n}{1+n} \\
\int_0^1 dx  (1-x) P^{(1,1)}_{m-1}(1-2x) P^{(1,1)}_{n-1}(1-2x) &=& \frac{\min\{m,n\}}{\max\{m,n\}+1}
\eea
for $m, n=1, 2,...$. Furthermore, due to Eq.~(\ref{Pi0}) and  Eq.~(\ref{Pi1}), the first sum on the RHS of Eq.~(\ref{integral1}) is equal to $(1-\Pi_0(t_2; p_0)-\Pi_1(t_1; p_0))/(p_0 (1-p_0))$ on using which we finally arrive at Eq.~(\ref{unequalsum}) in the main text. 
%

\section{Two time correlation function at intermediate times}
\label{app_two22}

Here, we consider the time regime $t_1 \ll 2 N_0$ and $2 N_0 \ll t_2-t_1 \ll 2 N$, and develop approximations for the sum on the RHS of Eq.~(\ref{unequalsum}) which can be written as $S_1+S_2$ where
\bea
S_1 &=& (1-p_0) \sum_{n=1}^\infty \frac{2 n+1}{n} P^{(1,1)}_{n-1}(1-2 p_0) e^{-\Lambda_{n} t_1} \sum_{m=1}^n (2 m+1) 
e^{-\Lambda_{m} (t_2-t_1) }  \label{app_S1a} \\
S_2 &=& (1-p_0) \sum_{n=1}^\infty (2 n+1) (n+1)P^{(1,1)}_{n-1}(1-2 p_0) e^{-\Lambda_{n} t_1} \sum_{m=n+1}^\infty \frac{2 m+1}{m (m+1)} e^{-\Lambda_{m} (t_2-t_1) } \label{S2full}
\eea
For $t_2-t_1 \ll 2 N$, as $m \gg 1$ contributes to $S_1$ and $S_2$, we approximate the sum over $m$ by an integral as the exact sums do not seem to be doable so that 
\bea
 \sum_{m=1}^n (2 m+1)
e^{-\Lambda_{m} (t_2-t_1) }  &{\approx}& \int_0^n dm \; 2 m 
e^{-\frac{m^2 (t_2-t_1)}{2 N} } =\frac{2 N \left(1-e^{-\frac{n^2 (t_2-t_1 )}{2 N}}\right)}{t_2-t_1} \\
\sum_{m=n+1}^\infty \frac{(2 m+1)}{m (m+1)} 
e^{-\Lambda_{m} (t_2-t_1) }&{\approx}& \int_n^\infty dm \; \frac{2}{m} e^{-\frac{m^2 (t_2-t_1)}{2 N} } =E_1 \left( \frac{n^2 (t_2-t_1 )}{2 N} \right)
\eea
Using the above approximations and Eq.~(\ref{PJapprox}), for $p_0 \to 0$, we then obtain
\bea
S_1 &\approx& 
\frac{2N^2}{N_0 (t_2-t_1)}  \left( e^{-\frac{2 N_0}{t_2}}-e^{-\frac{2 N_0}{t_1 }}\right) \approx \frac{2N  e^{-\frac{2 N_0}{t_2}}}{p_0 (t_2-t_1)} \\
S_2 &\approx& 
-\frac{1}{4 p_0^2} \int_0^{\sqrt{\frac{8 N_0}{t_2-t_1}}} dz \; z^2 J_1(z)  \times \ln \left(\frac{z^2 (t_2-t_1 )}{8 N_0} \right) \approx \frac{N^2}{(t_2-t_1)^2}
\eea
where, for $S_2$, we have used (6.2.4) of \cite{DLMF} for the power series expansion of the exponential integral. Then, using Eq.~(\ref{aHM}) and the above approximations in Eq.~(\ref{unequalsum}) we finally obtain
\bea
\Bav{\frac{1}{p(t_2) p(t_1)}}^* &\approx& \frac{1}{p_0}+\frac{2N  e^{-\frac{2 N_0}{t_2}}}{p_0 (t_2-t_1)}+ \frac{N^2}{(t_2-t_1)^2}- \frac{1}{p_0}  \frac{1- e^{-\frac{2 N_0}{t_2}}}{p_0} \approx \frac{N^2}{(t_2-t_1)^2} \label{corrint}
\eea

\section{Exponential decay of the two time correlation function }
\label{app_two2}

When $t_2-t_1 \gg 2 N$, it is sufficient to consider the lowest eigenvalue $\Lambda_m$ in the sum over $m$ in Eq.~(\ref{unequalsum}), but we need to consider the following three cases for the sum over $n$:

\noindent(i) For $t_1 \ll 2 N_0, t_2-t_1 \gg 2 N$, from Eq.~(\ref{unequalsum}), we can write 
\bea
\Bav{\frac{1}{p(t_2) p(t_1)}}^* &\approx& \frac{1}{p_0}+\sum_{n=1}^\infty  \frac{2 n+1}{n} P^{(1,1)}_{n-1}(1-2 p_0) \times 3
e^{-\frac{t_2-t_1}{N}-\Lambda_{n} t_1}  \\
&\approx&  \frac{1}{p_0}+ 6 e^{-\frac{t_2-t_1}{N}} \int_0^\infty dn e^{-\frac{n^2 t_1}{2 N}} \frac{J_1(2 n \sqrt{p_0})}{\sqrt{p_0}} \\
&\approx&  \frac{1}{p_0}+\frac{3}{p_0} e^{-\frac{t_2-t_1}{N}}
\eea
while $\bav{\frac{1}{p(t_1)}}^* \bav{\frac{1}{p(t_2)}}^* \approx p_0^{-1} (1+3 e^{-\frac{t_2}{N}})$ due to Eq.~(\ref{aHM}) so that
\bea
C(t_2, t_1)&\approx& \frac{3}{p_0} (e^{\frac{t_1}{N}}-1) e^{-\frac{t_2}{N}} \approx \frac{3 t_1}{N_0} e^{-\frac{t_2}{N}} \label{late1}
\eea

\noindent(ii) For $2 N_0 \ll t_1 \ll 2 N, t_2-t_1 \gg 2 N$, on proceeding as above, we obtain
\bea
C(t_2, t_1)&\approx& \frac{1- e^{-\frac{2 N_0}{t_1}}}{p_0}+\frac{3 (1-e^{-\frac{2 N_0}{t_1}}) e^{-\frac{t_2-t_1
   }{N}}}{p_0}- \frac{(1- e^{-\frac{2 N_0}{t_1}}) (1+3 e^{-\frac{t_2}{N}})}{p_0}  \\
   &\approx & \frac{3}{p_0} (1- e^{-\frac{2 N_0}{t_1}}) (e^{\frac{t_1}{N}}-1) e^{-\frac{t_2}{N}} \approx \frac{3 t_1}{N_0}  (1- e^{-\frac{2 N_0}{t_1}}) e^{-\frac{t_2}{N}}
\eea

\noindent(iii) For $t_1 \gg 2 N, t_2-t_1 \gg 2 N$, it is sufficient to keep terms corresponding to $n=m=1$ in the sum on the RHS of Eq.~(\ref{unequalsum}), and we obtain 
\bea
\Bav{\frac{1}{p(t_2) p(t_1)}}^* &\approx& 1+3 e^{-\frac{t_1}{N}}+ 9 e^{-\Lambda_{1} (t_2-t_1)-\Lambda_{1} t_1} 
\eea
and $\av{\frac{1}{p(t_2)}}^* \av{\frac{1}{p(t_1)}}^*=(1+3 e^{-\frac{t_2}{N}}) (1+3 e^{-\frac{t_1}{N}})$
so that
\bea
C(t_2, t_1)&\approx&6 e^{-\frac{t_2}{N}}
\eea

\section{Distribution of population size at the time of fixation}
\label{app_Nf}

Here, we discuss the distribution of population size at the time allele $1$ fixes in the uncorrelated model. 
As $\av{\Phi_1^{(u)}}^*$ given by  Eq.~(\ref{Pi1Fun}) is the cumulative fixation probability by time $t$, on taking its derivative with respect to time, 
we obtain
\bea
Q^{(u)}(N_f) &=&  x_0(1-x_0)  \sum_{n=1}^\infty   \frac{(-1)^{n+1} (2 n+1)}{n} P^{(1,1)}_{n-1}(1-2 x_0) e^{-\Lambda_n \int_0^t \frac{dt'}{p^*_{hm}(t')}}  \times \frac{\Lambda_n}{p^*_{hm}(t)} \\
&\approx& 3 x_0(1-x_0) e^{-\frac{1}{N} \int_0^t \frac{dt'}{p^*_{hm}(t')}}  \times \frac{1}{N p^*_{hm}(t)}
\eea
where, due to the discussion in Appendix~\ref{app_HM} for the fixation probability, we have retained only the term corresponding to the lowest eigenvalue. To change the variables from time to population size (which is a random variable), we approximate the relationship through the harmonic mean, $N_f = N p^*_{hm}(t)=\frac{N}{\av{1/p(t)}^*}$ where $p^*_{hm}(t)$ is given by Eq.~(\ref{phmdef}).  We then have
\bea
Q^{(u)}(N_f) &\approx& \frac{3 x_0(1-x_0)}{N_f} e^{- \frac{1}{N}\int_0^t dt' \bav{\frac{1}{p(t')}}^*}
\eea
For $t \ll 2 N_0$, from Eq.~(\ref{aHM1}), we find that  $N_f \approx N_0$ so that 
\bea
Q^{(u)}(N_f) &\approx& \frac{3 x_0(1-x_0)}{N_0} e^{-\frac{t}{N_0}} \approx  \frac{3 x_0(1-x_0)}{N_0} ~,~N_f \ll N_0 \label{Nf1}
\eea
whereas, for $2 N_0 \ll t \ll 2 N$, from Eq.~(\ref{aHM2}), we can write $N_f \approx \frac{t}{2}$ so that
\bea
Q^{(u)}(N_f) &\approx& \frac{3 x_0(1-x_0)}{N_f} e^{-\frac{1}{N} \int_{2 N_0}^t dt' \frac{2 N}{t'}} \\
&\approx&  \frac{3 x_0(1-x_0)}{N_f} \left(\frac{2 N_0}{t} \right)^2 \\
&=& \frac{3 x_0(1-x_0)}{N_f} \left(\frac{N_0}{N_f} \right)^2~,~N_f \gg N_0 \label{Nf2}
\eea




\begin{thebibliography}{59}%
\makeatletter
\providecommand \@ifxundefined [1]{%
 \@ifx{#1\undefined}
}%
\providecommand \@ifnum [1]{%
 \ifnum #1\expandafter \@firstoftwo
 \else \expandafter \@secondoftwo
 \fi
}%
\providecommand \@ifx [1]{%
 \ifx #1\expandafter \@firstoftwo
 \else \expandafter \@secondoftwo
 \fi
}%
\providecommand \natexlab [1]{#1}%
\providecommand \enquote  [1]{``#1''}%
\providecommand \bibnamefont  [1]{#1}%
\providecommand \bibfnamefont [1]{#1}%
\providecommand \citenamefont [1]{#1}%
\providecommand \href@noop [0]{\@secondoftwo}%
\providecommand \href [0]{\begingroup \@sanitize@url \@href}%
\providecommand \@href[1]{\@@startlink{#1}\@@href}%
\providecommand \@@href[1]{\endgroup#1\@@endlink}%
\providecommand \@sanitize@url [0]{\catcode `\\12\catcode `\$12\catcode
  `\&12\catcode `\#12\catcode `\^12\catcode `\_12\catcode `\%12\relax}%
\providecommand \@@startlink[1]{}%
\providecommand \@@endlink[0]{}%
\providecommand \url  [0]{\begingroup\@sanitize@url \@url }%
\providecommand \@url [1]{\endgroup\@href {#1}{\urlprefix }}%
\providecommand \urlprefix  [0]{URL }%
\providecommand \Eprint [0]{\href }%
\providecommand \doibase [0]{https://doi.org/}%
\providecommand \selectlanguage [0]{\@gobble}%
\providecommand \bibinfo  [0]{\@secondoftwo}%
\providecommand \bibfield  [0]{\@secondoftwo}%
\providecommand \translation [1]{[#1]}%
\providecommand \BibitemOpen [0]{}%
\providecommand \bibitemStop [0]{}%
\providecommand \bibitemNoStop [0]{.\EOS\space}%
\providecommand \EOS [0]{\spacefactor3000\relax}%
\providecommand \BibitemShut  [1]{\csname bibitem#1\endcsname}%
\let\auto@bib@innerbib\@empty
\bibitem [{\citenamefont {Darwin}(1859)}]{Darwin:1859}%
  \BibitemOpen
  \bibfield  {author} {\bibinfo {author} {\bibfnamefont {C.}~\bibnamefont
  {Darwin}},\ }\href@noop {} {\emph {\bibinfo {title} {The origin of species by
  means of natural selection}}}\ (\bibinfo  {publisher} {John Murray, London},\
  \bibinfo {year} {1859})\BibitemShut {NoStop}%
\bibitem [{\citenamefont {Charlesworth}\ and\ \citenamefont
  {Charlesworth}(2010)}]{Charlesworth:2010}%
  \BibitemOpen
  \bibfield  {author} {\bibinfo {author} {\bibfnamefont {B.}~\bibnamefont
  {Charlesworth}}\ and\ \bibinfo {author} {\bibfnamefont {D.}~\bibnamefont
  {Charlesworth}},\ }\href@noop {} {\emph {\bibinfo {title} {Elements of
  {E}volutionary {G}enetics}}}\ (\bibinfo  {publisher} {Roberts and Company
  Publishers, Greenwood Village CO},\ \bibinfo {year} {2010})\BibitemShut
  {NoStop}%
\bibitem [{\citenamefont {Ohta}\ and\ \citenamefont
  {Kojima}(1968)}]{Ohta:1968}%
  \BibitemOpen
  \bibfield  {author} {\bibinfo {author} {\bibfnamefont {T.}~\bibnamefont
  {Ohta}}\ and\ \bibinfo {author} {\bibfnamefont {K.-I.}\ \bibnamefont
  {Kojima}},\ }\bibfield  {title} {\bibinfo {title} {Survival probabilities of
  new inversions in large populations},\ }\href@noop {} {\bibfield  {journal}
  {\bibinfo  {journal} {Biometrics}\ }\textbf {\bibinfo {volume} {24}},\
  \bibinfo {pages} {501} (\bibinfo {year} {1968})}\BibitemShut {NoStop}%
\bibitem [{\citenamefont {Waxman}(2011)}]{Waxman:2011}%
  \BibitemOpen
  \bibfield  {author} {\bibinfo {author} {\bibfnamefont {D.}~\bibnamefont
  {Waxman}},\ }\bibfield  {title} {\bibinfo {title} {A unified treatment of the
  probability of fixation when population size and the strength of selection
  change over time},\ }\href@noop {} {\bibfield  {journal} {\bibinfo  {journal}
  {Genetics}\ }\textbf {\bibinfo {volume} {188}},\ \bibinfo {pages} {907}
  (\bibinfo {year} {2011})}\BibitemShut {NoStop}%
\bibitem [{\citenamefont {Uecker}\ and\ \citenamefont
  {Hermisson}(2011)}]{Uecker:2011}%
  \BibitemOpen
  \bibfield  {author} {\bibinfo {author} {\bibfnamefont {H.}~\bibnamefont
  {Uecker}}\ and\ \bibinfo {author} {\bibfnamefont {J.}~\bibnamefont
  {Hermisson}},\ }\bibfield  {title} {\bibinfo {title} {On the fixation process
  of a beneficial mutation in a variable environment},\ }\href@noop {}
  {\bibfield  {journal} {\bibinfo  {journal} {Genetics}\ }\textbf {\bibinfo
  {volume} {188}},\ \bibinfo {pages} {915} (\bibinfo {year}
  {2011})}\BibitemShut {NoStop}%
\bibitem [{\citenamefont {Devi}\ and\ \citenamefont {Jain}(2020)}]{Devi:2020}%
  \BibitemOpen
  \bibfield  {author} {\bibinfo {author} {\bibfnamefont {A.}~\bibnamefont
  {Devi}}\ and\ \bibinfo {author} {\bibfnamefont {K.}~\bibnamefont {Jain}},\
  }\bibfield  {title} {\bibinfo {title} {The impact of dominance on adaptation
  in changing environments},\ }\href@noop {} {\bibfield  {journal} {\bibinfo
  {journal} {Genetics}\ }\textbf {\bibinfo {volume} {216}},\ \bibinfo {pages}
  {227} (\bibinfo {year} {2020})}\BibitemShut {NoStop}%
\bibitem [{\citenamefont {Assaf}\ \emph {et~al.}(2008)\citenamefont {Assaf},
  \citenamefont {Kamenev},\ and\ \citenamefont {Meerson}}]{Assaf:2008}%
  \BibitemOpen
  \bibfield  {author} {\bibinfo {author} {\bibfnamefont {M.}~\bibnamefont
  {Assaf}}, \bibinfo {author} {\bibfnamefont {A.}~\bibnamefont {Kamenev}},\
  and\ \bibinfo {author} {\bibfnamefont {B.}~\bibnamefont {Meerson}},\
  }\bibfield  {title} {\bibinfo {title} {Population extinction in a
  time-modulated environment},\ }\href@noop {} {\bibfield  {journal} {\bibinfo
  {journal} {Phys. Rev. E}\ }\textbf {\bibinfo {volume} {78}},\ \bibinfo
  {pages} {041123} (\bibinfo {year} {2008})}\BibitemShut {NoStop}%
\bibitem [{\citenamefont {Kaushik}\ and\ \citenamefont
  {Jain}(2021)}]{Kaushik:2021}%
  \BibitemOpen
  \bibfield  {author} {\bibinfo {author} {\bibfnamefont {S.}~\bibnamefont
  {Kaushik}}\ and\ \bibinfo {author} {\bibfnamefont {K.}~\bibnamefont {Jain}},\
  }\bibfield  {title} {\bibinfo {title} {Time to fixation in changing
  environments},\ }\href@noop {} {\bibfield  {journal} {\bibinfo  {journal}
  {Genetics}\ }\textbf {\bibinfo {volume} {219}},\ \bibinfo {pages} {iyab 148}
  (\bibinfo {year} {2021})}\BibitemShut {NoStop}%
\bibitem [{\citenamefont {Jain}\ and\ \citenamefont
  {Kaushik}(2022)}]{Jain:2022}%
  \BibitemOpen
  \bibfield  {author} {\bibinfo {author} {\bibfnamefont {K.}~\bibnamefont
  {Jain}}\ and\ \bibinfo {author} {\bibfnamefont {S.}~\bibnamefont {Kaushik}},\
  }\bibfield  {title} {\bibinfo {title} {Joint effect of changing selection and
  demography on the site frequency spectrum},\ }\href@noop {} {\bibfield
  {journal} {\bibinfo  {journal} {Theor Pop Biol}\ }\textbf {\bibinfo {volume}
  {146}},\ \bibinfo {pages} {46} (\bibinfo {year} {2022})}\BibitemShut
  {NoStop}%
\bibitem [{\citenamefont {Kaushik}(2023)}]{Kaushik:2023}%
  \BibitemOpen
  \bibfield  {author} {\bibinfo {author} {\bibfnamefont {S.}~\bibnamefont
  {Kaushik}},\ }\bibfield  {title} {\bibinfo {title} {Effect of beneficial
  sweeps and background selection on genetic diversity in changing
  environments},\ }\href@noop {} {\bibfield  {journal} {\bibinfo  {journal} {J
  Theor Biol.}\ }\textbf {\bibinfo {volume} {562}},\ \bibinfo {pages} {111431}
  (\bibinfo {year} {2023})}\BibitemShut {NoStop}%
\bibitem [{\citenamefont {Nei}\ \emph {et~al.}(1975)\citenamefont {Nei},
  \citenamefont {Maruyama},\ and\ \citenamefont {Chakraborty}}]{Nei:1975}%
  \BibitemOpen
  \bibfield  {author} {\bibinfo {author} {\bibfnamefont {M.}~\bibnamefont
  {Nei}}, \bibinfo {author} {\bibfnamefont {T.}~\bibnamefont {Maruyama}},\ and\
  \bibinfo {author} {\bibfnamefont {R.}~\bibnamefont {Chakraborty}},\
  }\bibfield  {title} {\bibinfo {title} {The bottleneck effect and genetic
  variability in populations},\ }\href@noop {} {\bibfield  {journal} {\bibinfo
  {journal} {Evolution}\ }\textbf {\bibinfo {volume} {29}},\ \bibinfo {pages}
  {1} (\bibinfo {year} {1975})}\BibitemShut {NoStop}%
\bibitem [{\citenamefont {Maruyama}\ and\ \citenamefont
  {Fuerst}(1985)}]{Maruyama:1985c}%
  \BibitemOpen
  \bibfield  {author} {\bibinfo {author} {\bibfnamefont {T.}~\bibnamefont
  {Maruyama}}\ and\ \bibinfo {author} {\bibfnamefont {P.~A.}\ \bibnamefont
  {Fuerst}},\ }\bibfield  {title} {\bibinfo {title} {Population bottlenecks and
  nonequilibrium models in population genetics. iii. genic homozygosity in
  populations which experience periodic bottlenecks},\ }\href@noop {}
  {\bibfield  {journal} {\bibinfo  {journal} {Genetics}\ }\textbf {\bibinfo
  {volume} {111}},\ \bibinfo {pages} {691} (\bibinfo {year}
  {1985})}\BibitemShut {NoStop}%
\bibitem [{\citenamefont {Williamson}\ \emph {et~al.}(2005)\citenamefont
  {Williamson}, \citenamefont {Hernandez}, \citenamefont {Fledel-Alon},
  \citenamefont {Zhu}, \citenamefont {Nielsen},\ and\ \citenamefont
  {Bustamante}}]{Williamson:2005}%
  \BibitemOpen
  \bibfield  {author} {\bibinfo {author} {\bibfnamefont {S.~H.}\ \bibnamefont
  {Williamson}}, \bibinfo {author} {\bibfnamefont {R.}~\bibnamefont
  {Hernandez}}, \bibinfo {author} {\bibfnamefont {A.}~\bibnamefont
  {Fledel-Alon}}, \bibinfo {author} {\bibfnamefont {L.}~\bibnamefont {Zhu}},
  \bibinfo {author} {\bibfnamefont {R.}~\bibnamefont {Nielsen}},\ and\ \bibinfo
  {author} {\bibfnamefont {C.~D.}\ \bibnamefont {Bustamante}},\ }\bibfield
  {title} {\bibinfo {title} {Simultaneous inference of selection and population
  growth from patterns of variation in the human genome},\ }\href@noop {}
  {\bibfield  {journal} {\bibinfo  {journal} {Proc. Nat Acad. Sci. USA}\
  }\textbf {\bibinfo {volume} {102}},\ \bibinfo {pages} {7882–7887} (\bibinfo
  {year} {2005})}\BibitemShut {NoStop}%
\bibitem [{\citenamefont {Waxman}(2012)}]{Waxman:2012}%
  \BibitemOpen
  \bibfield  {author} {\bibinfo {author} {\bibfnamefont {D.}~\bibnamefont
  {Waxman}},\ }\bibfield  {title} {\bibinfo {title} {Population growth enhances
  the mean fixation time of neutral mutations and the persistence of neutral
  variation},\ }\href@noop {} {\bibfield  {journal} {\bibinfo  {journal}
  {Genetics}\ }\textbf {\bibinfo {volume} {191}},\ \bibinfo {pages} {561}
  (\bibinfo {year} {2012})}\BibitemShut {NoStop}%
\bibitem [{\citenamefont {Nakamura}\ \emph {et~al.}(2018)\citenamefont
  {Nakamura}, \citenamefont {Teshima},\ and\ \citenamefont
  {Tachida}}]{Nakamura:2018}%
  \BibitemOpen
  \bibfield  {author} {\bibinfo {author} {\bibfnamefont {H.}~\bibnamefont
  {Nakamura}}, \bibinfo {author} {\bibfnamefont {K.}~\bibnamefont {Teshima}},\
  and\ \bibinfo {author} {\bibfnamefont {H.}~\bibnamefont {Tachida}},\
  }\bibfield  {title} {\bibinfo {title} {Effects of cyclic changes in
  population size on neutral genetic diversity},\ }\href@noop {} {\bibfield
  {journal} {\bibinfo  {journal} {Ecology and Evolution}\ }\textbf {\bibinfo
  {volume} {8}},\ \bibinfo {pages} {9362} (\bibinfo {year} {2018})}\BibitemShut
  {NoStop}%
\bibitem [{\citenamefont {Kaushik}\ \emph {et~al.}(2025)\citenamefont
  {Kaushik}, \citenamefont {Jain},\ and\ \citenamefont {Johri}}]{Kaushik:2025}%
  \BibitemOpen
  \bibfield  {author} {\bibinfo {author} {\bibfnamefont {S.}~\bibnamefont
  {Kaushik}}, \bibinfo {author} {\bibfnamefont {K.}~\bibnamefont {Jain}},\ and\
  \bibinfo {author} {\bibfnamefont {P.}~\bibnamefont {Johri}},\ }\bibfield
  {title} {\bibinfo {title} {Genetic diversity during selective sweeps in
  non-recombining populations},\ }\href
  {https://doi.org/{10.1101/2024.09.12.612756}} {\bibfield  {journal} {\bibinfo
   {journal} {Genetics}\ }\textbf {\bibinfo {volume} {xxx}},\ \bibinfo {pages}
  {xxx} (\bibinfo {year} {2025})}\BibitemShut {NoStop}%
\bibitem [{\citenamefont {Takahata}\ \emph {et~al.}(1975)\citenamefont
  {Takahata}, \citenamefont {Kazushige},\ and\ \citenamefont
  {Matsuda}}]{Takahata:1975}%
  \BibitemOpen
  \bibfield  {author} {\bibinfo {author} {\bibfnamefont {N.}~\bibnamefont
  {Takahata}}, \bibinfo {author} {\bibfnamefont {I.}~\bibnamefont
  {Kazushige}},\ and\ \bibinfo {author} {\bibfnamefont {H.}~\bibnamefont
  {Matsuda}},\ }\bibfield  {title} {\bibinfo {title} {Effect of temporal
  fluctuation of selection coefficient on gene frequency in a population},\
  }\href@noop {} {\bibfield  {journal} {\bibinfo  {journal} {Proc. Nat. Acad.
  Sci. USA}\ }\textbf {\bibinfo {volume} {72}},\ \bibinfo {pages} {4541}
  (\bibinfo {year} {1975})}\BibitemShut {NoStop}%
\bibitem [{\citenamefont {Gillespie}(1991)}]{Gillespie:1991}%
  \BibitemOpen
  \bibfield  {author} {\bibinfo {author} {\bibfnamefont {J.~H.}\ \bibnamefont
  {Gillespie}},\ }\href@noop {} {\emph {\bibinfo {title} {The Causes of
  Molecular Evolution}}}\ (\bibinfo  {publisher} {Oxford University Press,
  Oxford},\ \bibinfo {year} {1991})\BibitemShut {NoStop}%
\bibitem [{\citenamefont {Huerta-Sanchez}\ \emph {et~al.}(2008)\citenamefont
  {Huerta-Sanchez}, \citenamefont {Durrett},\ and\ \citenamefont
  {Bustamante}}]{HuertaSanchez:2008}%
  \BibitemOpen
  \bibfield  {author} {\bibinfo {author} {\bibfnamefont {E.}~\bibnamefont
  {Huerta-Sanchez}}, \bibinfo {author} {\bibfnamefont {R.}~\bibnamefont
  {Durrett}},\ and\ \bibinfo {author} {\bibfnamefont {C.~D.}\ \bibnamefont
  {Bustamante}},\ }\bibfield  {title} {\bibinfo {title} {Population genetics of
  polymorphism and divergence under fluctuating selection},\ }\href@noop {}
  {\bibfield  {journal} {\bibinfo  {journal} {Genetics}\ }\textbf {\bibinfo
  {volume} {178}},\ \bibinfo {pages} {325} (\bibinfo {year}
  {2008})}\BibitemShut {NoStop}%
\bibitem [{\citenamefont {Sj{\"o}din}\ \emph {et~al.}(2005)\citenamefont
  {Sj{\"o}din}, \citenamefont {Kaj}, \citenamefont {Krone}, \citenamefont
  {Lascoux},\ and\ \citenamefont {Nordborg}}]{Sjodin:2005}%
  \BibitemOpen
  \bibfield  {author} {\bibinfo {author} {\bibfnamefont {P.}~\bibnamefont
  {Sj{\"o}din}}, \bibinfo {author} {\bibfnamefont {I.}~\bibnamefont {Kaj}},
  \bibinfo {author} {\bibfnamefont {S.}~\bibnamefont {Krone}}, \bibinfo
  {author} {\bibfnamefont {M.}~\bibnamefont {Lascoux}},\ and\ \bibinfo {author}
  {\bibfnamefont {M.}~\bibnamefont {Nordborg}},\ }\bibfield  {title} {\bibinfo
  {title} {On the meaning and existence of effective population size},\
  }\href@noop {} {\bibfield  {journal} {\bibinfo  {journal} {Genetics}\
  }\textbf {\bibinfo {volume} {169}},\ \bibinfo {pages} {1061} (\bibinfo {year}
  {2005})}\BibitemShut {NoStop}%
\bibitem [{\citenamefont {Schaper}\ \emph {et~al.}(2012)\citenamefont
  {Schaper}, \citenamefont {Eriksson}, \citenamefont {Rafajlovic},
  \citenamefont {Sagitov},\ and\ \citenamefont {Mehlig}}]{Schaper:2012}%
  \BibitemOpen
  \bibfield  {author} {\bibinfo {author} {\bibfnamefont {E.}~\bibnamefont
  {Schaper}}, \bibinfo {author} {\bibfnamefont {A.}~\bibnamefont {Eriksson}},
  \bibinfo {author} {\bibfnamefont {M.}~\bibnamefont {Rafajlovic}}, \bibinfo
  {author} {\bibfnamefont {S.}~\bibnamefont {Sagitov}},\ and\ \bibinfo {author}
  {\bibfnamefont {B.}~\bibnamefont {Mehlig}},\ }\bibfield  {title} {\bibinfo
  {title} {Linkage disequilibrium under recurrent bottlenecks},\ }\href@noop {}
  {\bibfield  {journal} {\bibinfo  {journal} {Genetics}\ }\textbf {\bibinfo
  {volume} {190}},\ \bibinfo {pages} {217} (\bibinfo {year}
  {2012})}\BibitemShut {NoStop}%
\bibitem [{\citenamefont {Živković}\ \emph {et~al.}(2015)\citenamefont
  {Živković}, \citenamefont {Steinrücken}, \citenamefont {Song},\ and\
  \citenamefont {Stephan}}]{Zivkovic:2015}%
  \BibitemOpen
  \bibfield  {author} {\bibinfo {author} {\bibfnamefont {D.}~\bibnamefont
  {Živković}}, \bibinfo {author} {\bibfnamefont {M.}~\bibnamefont
  {Steinrücken}}, \bibinfo {author} {\bibfnamefont {Y.~S.}\ \bibnamefont
  {Song}},\ and\ \bibinfo {author} {\bibfnamefont {W.}~\bibnamefont
  {Stephan}},\ }\bibfield  {title} {\bibinfo {title} {Transition densities and
  sample frequency spectra of diffusion processes with selection and variable
  population size},\ }\href@noop {} {\bibfield  {journal} {\bibinfo  {journal}
  {Genetics}\ }\textbf {\bibinfo {volume} {200}},\ \bibinfo {pages} {601}
  (\bibinfo {year} {2015})}\BibitemShut {NoStop}%
\bibitem [{\citenamefont {Asker}\ \emph {et~al.}()\citenamefont {Asker},
  \citenamefont {Swailem}, \citenamefont {T{\"a}uber},\ and\ \citenamefont
  {Mobilia}}]{Asker:2025}%
  \BibitemOpen
  \bibfield  {author} {\bibinfo {author} {\bibfnamefont {M.}~\bibnamefont
  {Asker}}, \bibinfo {author} {\bibfnamefont {M.}~\bibnamefont {Swailem}},
  \bibinfo {author} {\bibfnamefont {U.~C.}\ \bibnamefont {T{\"a}uber}},\ and\
  \bibinfo {author} {\bibfnamefont {M.}~\bibnamefont {Mobilia}},\ }\bibfield
  {title} {\bibinfo {title} {Fixation and extinction in time-fluctuating
  spatially structured metapopulations},\ }\href@noop {} {\bibinfo  {journal}
  {arXiv.2504.08433}\ }\BibitemShut {NoStop}%
\bibitem [{\citenamefont {Jain}(2025)}]{Jain:2025}%
  \BibitemOpen
\bibfield  {journal} {  }\bibfield  {author} {\bibinfo {author} {\bibfnamefont
  {K.}~\bibnamefont {Jain}},\ }\bibfield  {title} {\bibinfo {title}
  {Evolutionary dynamics under demographic fluctuations: beyond the effective
  population size},\ }\href
  {https://doi.org/https://doi.org/10.1101/2025.04.12.648494} {\bibfield
  {journal} {\bibinfo  {journal} {bioRxiv 2025.04.12.648494}\ }\textbf
  {\bibinfo {volume} {-}},\  (\bibinfo {year} {2025})}\BibitemShut {NoStop}%
\bibitem [{\citenamefont {Meyer}\ \emph {et~al.}(2024)\citenamefont {Meyer},
  \citenamefont {Taitelbaum}, \citenamefont {Assaf},\ and\ \citenamefont
  {Shnerb}}]{Meyer:2024}%
  \BibitemOpen
  \bibfield  {author} {\bibinfo {author} {\bibfnamefont {I.}~\bibnamefont
  {Meyer}}, \bibinfo {author} {\bibfnamefont {A.}~\bibnamefont {Taitelbaum}},
  \bibinfo {author} {\bibfnamefont {M.}~\bibnamefont {Assaf}},\ and\ \bibinfo
  {author} {\bibfnamefont {N.~M.}\ \bibnamefont {Shnerb}},\ }\bibfield  {title}
  {\bibinfo {title} {Population dynamics in a time-varying environment with
  fat-tailed correlations},\ }\href@noop {} {\bibfield  {journal} {\bibinfo
  {journal} {Phys Rev E}\ }\textbf {\bibinfo {volume} {110}},\ \bibinfo {pages}
  {L012401} (\bibinfo {year} {2024})}\BibitemShut {NoStop}%
\bibitem [{\citenamefont {Lambert}(2006)}]{Lambert:2006}%
  \BibitemOpen
  \bibfield  {author} {\bibinfo {author} {\bibfnamefont {A.}~\bibnamefont
  {Lambert}},\ }\bibfield  {title} {\bibinfo {title} {Probability of fixation
  under weak selection: A branching process unifying approach},\ }\href@noop {}
  {\bibfield  {journal} {\bibinfo  {journal} {Theor Popul Biol.}\ }\textbf
  {\bibinfo {volume} {69}},\ \bibinfo {pages} {419} (\bibinfo {year}
  {2006})}\BibitemShut {NoStop}%
\bibitem [{\citenamefont {Parsons}\ and\ \citenamefont
  {Quince}(2007{\natexlab{a}})}]{Parsons:2007a}%
  \BibitemOpen
  \bibfield  {author} {\bibinfo {author} {\bibfnamefont {T.~L.}\ \bibnamefont
  {Parsons}}\ and\ \bibinfo {author} {\bibfnamefont {C.}~\bibnamefont
  {Quince}},\ }\bibfield  {title} {\bibinfo {title} {Fixation in haploid
  populations exhibiting density dependence i: The non-neutral case},\
  }\href@noop {} {\bibfield  {journal} {\bibinfo  {journal} {Theor Popul
  Biol.}\ }\textbf {\bibinfo {volume} {72}},\ \bibinfo {pages} {121} (\bibinfo
  {year} {2007}{\natexlab{a}})}\BibitemShut {NoStop}%
\bibitem [{\citenamefont {Parsons}\ and\ \citenamefont
  {Quince}(2007{\natexlab{b}})}]{Parsons:2007b}%
  \BibitemOpen
  \bibfield  {author} {\bibinfo {author} {\bibfnamefont {T.~L.}\ \bibnamefont
  {Parsons}}\ and\ \bibinfo {author} {\bibfnamefont {C.}~\bibnamefont
  {Quince}},\ }\bibfield  {title} {\bibinfo {title} {Fixation in haploid
  populations exhibiting density dependence ii: the quasi-neutral case},\
  }\href@noop {} {\bibfield  {journal} {\bibinfo  {journal} {Theor Popul
  Biol.}\ }\textbf {\bibinfo {volume} {72}},\ \bibinfo {pages} {468} (\bibinfo
  {year} {2007}{\natexlab{b}})}\BibitemShut {NoStop}%
\bibitem [{\citenamefont {Parsons}\ \emph {et~al.}(2010)\citenamefont
  {Parsons}, \citenamefont {Quince},\ and\ \citenamefont
  {Plotkin}}]{Parsons:2010}%
  \BibitemOpen
  \bibfield  {author} {\bibinfo {author} {\bibfnamefont {T.~L.}\ \bibnamefont
  {Parsons}}, \bibinfo {author} {\bibfnamefont {C.}~\bibnamefont {Quince}},\
  and\ \bibinfo {author} {\bibfnamefont {J.~B.}\ \bibnamefont {Plotkin}},\
  }\bibfield  {title} {\bibinfo {title} {Some consequences of demographic
  stochasticity in population genetics},\ }\href@noop {} {\bibfield  {journal}
  {\bibinfo  {journal} {Genetics}\ }\textbf {\bibinfo {volume} {185}},\
  \bibinfo {pages} {1345} (\bibinfo {year} {2010})}\BibitemShut {NoStop}%
\bibitem [{\citenamefont {Engen}\ \emph {et~al.}(2009)\citenamefont {Engen},
  \citenamefont {Lande},\ and\ \citenamefont {Saether}}]{Engen:2009}%
  \BibitemOpen
  \bibfield  {author} {\bibinfo {author} {\bibfnamefont {S.}~\bibnamefont
  {Engen}}, \bibinfo {author} {\bibfnamefont {R.}~\bibnamefont {Lande}},\ and\
  \bibinfo {author} {\bibfnamefont {B.-E.}\ \bibnamefont {Saether}},\
  }\bibfield  {title} {\bibinfo {title} {Fixation probability of beneficial
  mutations in a fluctuating population},\ }\href@noop {} {\bibfield  {journal}
  {\bibinfo  {journal} {Genetics Research}\ }\textbf {\bibinfo {volume}
  {91(1)}},\ \bibinfo {pages} {73} (\bibinfo {year} {2009})}\BibitemShut
  {NoStop}%
\bibitem [{\citenamefont {Czuppon}\ and\ \citenamefont
  {Traulsen}(2018)}]{Czuppon:2018}%
  \BibitemOpen
  \bibfield  {author} {\bibinfo {author} {\bibfnamefont {P.}~\bibnamefont
  {Czuppon}}\ and\ \bibinfo {author} {\bibfnamefont {A.}~\bibnamefont
  {Traulsen}},\ }\bibfield  {title} {\bibinfo {title} {Fixation probabilities
  in populations under demographic fluctuations},\ }\href@noop {} {\bibfield
  {journal} {\bibinfo  {journal} {J Math Biol.}\ }\textbf {\bibinfo {volume}
  {77}},\ \bibinfo {pages} {1233} (\bibinfo {year} {2018})}\BibitemShut
  {NoStop}%
\bibitem [{\citenamefont {Kendall}(1960)}]{Kendall:1960}%
  \BibitemOpen
  \bibfield  {author} {\bibinfo {author} {\bibfnamefont {D.~G.}\ \bibnamefont
  {Kendall}},\ }\bibfield  {title} {\bibinfo {title} {Birth-and-death
  processes, and the theory of carcinogenesis},\ }\href@noop {} {\bibfield
  {journal} {\bibinfo  {journal} {Biometrika}\ }\textbf {\bibinfo {volume}
  {47}},\ \bibinfo {pages} {13} (\bibinfo {year} {1960})}\BibitemShut {NoStop}%
\bibitem [{\citenamefont {Durrett}(2013)}]{Durrett:2013}%
  \BibitemOpen
  \bibfield  {author} {\bibinfo {author} {\bibfnamefont {R.}~\bibnamefont
  {Durrett}},\ }\bibfield  {title} {\bibinfo {title} {Population genetics of
  neutral mutations in exponentially growing cancer cell populations},\
  }\href@noop {} {\bibfield  {journal} {\bibinfo  {journal} {Ann. Appl.
  Probab.}\ }\textbf {\bibinfo {volume} {23}},\ \bibinfo {pages} {230}
  (\bibinfo {year} {2013})}\BibitemShut {NoStop}%
\bibitem [{\citenamefont {Kessler}\ and\ \citenamefont
  {Levine}(2013)}]{Kessler:2013}%
  \BibitemOpen
  \bibfield  {author} {\bibinfo {author} {\bibfnamefont {D.~A.}\ \bibnamefont
  {Kessler}}\ and\ \bibinfo {author} {\bibfnamefont {H.}~\bibnamefont
  {Levine}},\ }\bibfield  {title} {\bibinfo {title} {Large population solution
  of the stochastic {L}uria-{D}elbr{\"u}ck evolution model},\ }\href@noop {}
  {\bibfield  {journal} {\bibinfo  {journal} {Proc. Nat. Acad. Sci. USA}\
  }\textbf {\bibinfo {volume} {110}},\ \bibinfo {pages} {11682–11687}
  (\bibinfo {year} {2013})}\BibitemShut {NoStop}%
\bibitem [{\citenamefont {Durrett}(2015)}]{Durrett:2015}%
  \BibitemOpen
  \bibfield  {author} {\bibinfo {author} {\bibfnamefont {R.}~\bibnamefont
  {Durrett}},\ }\href@noop {} {\emph {\bibinfo {title} {Branching process
  models of cancer}}}\ (\bibinfo  {publisher} {Springer, New York},\ \bibinfo
  {year} {2015})\BibitemShut {NoStop}%
\bibitem [{\citenamefont {Ohtsuki}\ and\ \citenamefont
  {Innan}(2017)}]{Ohtsuki:2017}%
  \BibitemOpen
  \bibfield  {author} {\bibinfo {author} {\bibfnamefont {H.}~\bibnamefont
  {Ohtsuki}}\ and\ \bibinfo {author} {\bibfnamefont {H.}~\bibnamefont
  {Innan}},\ }\bibfield  {title} {\bibinfo {title} {Forward and backward
  evolutionary processes and allele frequency spectrum in a cancer cell
  population},\ }\href@noop {} {\bibfield  {journal} {\bibinfo  {journal}
  {Theor Pop. Biol.}\ }\textbf {\bibinfo {volume} {117}},\ \bibinfo {pages}
  {43} (\bibinfo {year} {2017})}\BibitemShut {NoStop}%
\bibitem [{\citenamefont {Cheek}\ and\ \citenamefont
  {Antal}(2018)}]{Cheek:2018}%
  \BibitemOpen
  \bibfield  {author} {\bibinfo {author} {\bibfnamefont {D.}~\bibnamefont
  {Cheek}}\ and\ \bibinfo {author} {\bibfnamefont {T.}~\bibnamefont {Antal}},\
  }\bibfield  {title} {\bibinfo {title} {Mutation frequencies in a
  birth–death branching process},\ }\href@noop {} {\bibfield  {journal}
  {\bibinfo  {journal} {Ann. Appl. Probab.}\ }\textbf {\bibinfo {volume}
  {28}},\ \bibinfo {pages} {3922} (\bibinfo {year} {2018})}\BibitemShut
  {NoStop}%
\bibitem [{\citenamefont {Gunnarsson}\ \emph {et~al.}(2021)\citenamefont
  {Gunnarsson}, \citenamefont {Leder},\ and\ \citenamefont
  {Foo}}]{Gunnarsson:2021}%
  \BibitemOpen
  \bibfield  {author} {\bibinfo {author} {\bibfnamefont {E.~B.}\ \bibnamefont
  {Gunnarsson}}, \bibinfo {author} {\bibfnamefont {K.}~\bibnamefont {Leder}},\
  and\ \bibinfo {author} {\bibfnamefont {J.}~\bibnamefont {Foo}},\ }\bibfield
  {title} {\bibinfo {title} {Exact site frequency spectra of neutrally evolving
  tumors: A transition between power laws reveals a signature of cell
  viability},\ }\href@noop {} {\bibfield  {journal} {\bibinfo  {journal} {Theor
  Popul Biol}\ }\textbf {\bibinfo {volume} {142}},\ \bibinfo {pages} {67}
  (\bibinfo {year} {2021})}\BibitemShut {NoStop}%
\bibitem [{\citenamefont {Nicholson}\ \emph {et~al.}(2023)\citenamefont
  {Nicholson}, \citenamefont {Cheek},\ and\ \citenamefont
  {Antal}}]{Nicholson:2023}%
  \BibitemOpen
  \bibfield  {author} {\bibinfo {author} {\bibfnamefont {M.~D.}\ \bibnamefont
  {Nicholson}}, \bibinfo {author} {\bibfnamefont {D.}~\bibnamefont {Cheek}},\
  and\ \bibinfo {author} {\bibfnamefont {T.}~\bibnamefont {Antal}},\ }\bibfield
   {title} {\bibinfo {title} {Sequential mutations in exponentially growing
  populations},\ }\href@noop {} {\bibfield  {journal} {\bibinfo  {journal}
  {PLoS Comput Biol 19(7)}\ }\textbf {\bibinfo {volume} {19}},\ \bibinfo
  {pages} {e1011289} (\bibinfo {year} {2023})}\BibitemShut {NoStop}%
\bibitem [{\citenamefont {Mafessoni}\ and\ \citenamefont
  {Lachmann}(2015)}]{Mafessoni:2015}%
  \BibitemOpen
  \bibfield  {author} {\bibinfo {author} {\bibfnamefont {F.}~\bibnamefont
  {Mafessoni}}\ and\ \bibinfo {author} {\bibfnamefont {M.}~\bibnamefont
  {Lachmann}},\ }\bibfield  {title} {\bibinfo {title} {Selective strolls:
  fixation and extinction in diploids are slower for weakly selected mutations
  than for neutral ones},\ }\href@noop {} {\bibfield  {journal} {\bibinfo
  {journal} {Genetics}\ }\textbf {\bibinfo {volume} {201}},\ \bibinfo {pages}
  {1581} (\bibinfo {year} {2015})}\BibitemShut {NoStop}%
\bibitem [{\citenamefont {Moinet}\ \emph {et~al.}(2022)\citenamefont {Moinet},
  \citenamefont {Schlichta}, \citenamefont {Peischl},\ and\ \citenamefont
  {Excoffier}}]{Moinet:2022}%
  \BibitemOpen
  \bibfield  {author} {\bibinfo {author} {\bibfnamefont {A.}~\bibnamefont
  {Moinet}}, \bibinfo {author} {\bibfnamefont {F.}~\bibnamefont {Schlichta}},
  \bibinfo {author} {\bibfnamefont {S.}~\bibnamefont {Peischl}},\ and\ \bibinfo
  {author} {\bibfnamefont {L.}~\bibnamefont {Excoffier}},\ }\bibfield  {title}
  {\bibinfo {title} {Strong neutral sweeps occurring during a population
  contraction},\ }\href@noop {} {\bibfield  {journal} {\bibinfo  {journal}
  {Genetics}\ }\textbf {\bibinfo {volume} {220}},\ \bibinfo {pages} {iyac021}
  (\bibinfo {year} {2022})}\BibitemShut {NoStop}%
\bibitem [{\citenamefont {Maynard~Smith}\ and\ \citenamefont
  {Haigh}(1974)}]{Smith:1974}%
  \BibitemOpen
  \bibfield  {author} {\bibinfo {author} {\bibfnamefont {J.}~\bibnamefont
  {Maynard~Smith}}\ and\ \bibinfo {author} {\bibfnamefont {J.}~\bibnamefont
  {Haigh}},\ }\bibfield  {title} {\bibinfo {title} {Hitchhiking effect of a
  favourable gene},\ }\href@noop {} {\bibfield  {journal} {\bibinfo  {journal}
  {Genet. Res.}\ }\textbf {\bibinfo {volume} {23}},\ \bibinfo {pages} {23}
  (\bibinfo {year} {1974})}\BibitemShut {NoStop}%
\bibitem [{\citenamefont {Charlesworth}\ and\ \citenamefont
  {Jensen}(2021)}]{Charlesworth:2021}%
  \BibitemOpen
  \bibfield  {author} {\bibinfo {author} {\bibfnamefont {B.}~\bibnamefont
  {Charlesworth}}\ and\ \bibinfo {author} {\bibfnamefont {J.~D.}\ \bibnamefont
  {Jensen}},\ }\bibfield  {title} {\bibinfo {title} {Effects of selection at
  linked sites on patterns of genetic variability},\ }\href@noop {} {\bibfield
  {journal} {\bibinfo  {journal} {Annu Rev Ecol Evol Syst.}\ }\textbf {\bibinfo
  {volume} {52}},\ \bibinfo {pages} {177} (\bibinfo {year} {2021})}\BibitemShut
  {NoStop}%
\bibitem [{\citenamefont {Lewontin}(1974)}]{Lewontin:1974}%
  \BibitemOpen
  \bibfield  {author} {\bibinfo {author} {\bibfnamefont {R.~C.}\ \bibnamefont
  {Lewontin}},\ }\href@noop {} {\emph {\bibinfo {title} {The Genetic Basis of
  Evolutionary Change}}}\ (\bibinfo  {publisher} {New York: Columbia University
  Press},\ \bibinfo {year} {1974})\BibitemShut {NoStop}%
\bibitem [{\citenamefont {Buffalo}(2021)}]{Buffalo:2021}%
  \BibitemOpen
  \bibfield  {author} {\bibinfo {author} {\bibfnamefont {V.}~\bibnamefont
  {Buffalo}},\ }\bibfield  {title} {\bibinfo {title} {Quantifying the
  relationship between genetic diversity and population size suggests natural
  selection cannot explain {L}ewontin’s paradox},\ }\href@noop {} {\bibfield
  {journal} {\bibinfo  {journal} {eLife}\ }\textbf {\bibinfo {volume} {10}},\
  \bibinfo {pages} {e67509} (\bibinfo {year} {2021})}\BibitemShut {NoStop}%
\bibitem [{\citenamefont {Kimura}(1983)}]{Kimura:1983}%
  \BibitemOpen
  \bibfield  {author} {\bibinfo {author} {\bibfnamefont {M.}~\bibnamefont
  {Kimura}},\ }\href@noop {} {\emph {\bibinfo {title} {The Neutral Theory of
  Molecular Evolution}}}\ (\bibinfo  {publisher} {Cambridge University Press},\
  \bibinfo {year} {1983})\BibitemShut {NoStop}%
\bibitem [{\citenamefont {Terhorst}\ \emph {et~al.}(2015)\citenamefont
  {Terhorst}, \citenamefont {C.},\ and\ \citenamefont {Song}}]{Terhorst:2015}%
  \BibitemOpen
  \bibfield  {author} {\bibinfo {author} {\bibfnamefont {J.}~\bibnamefont
  {Terhorst}}, \bibinfo {author} {\bibfnamefont {S.}~\bibnamefont {C.}},\ and\
  \bibinfo {author} {\bibfnamefont {Y.~S.}\ \bibnamefont {Song}},\ }\bibfield
  {title} {\bibinfo {title} {Multi-locus analysis of genomic time series data
  from experimental evolution},\ }\href@noop {} {\bibfield  {journal} {\bibinfo
   {journal} {PLoS Genet}\ }\textbf {\bibinfo {volume} {11}},\ \bibinfo {pages}
  {e1005069} (\bibinfo {year} {2015})}\BibitemShut {NoStop}%
\bibitem [{\citenamefont {Blomberg}\ \emph {et~al.}(2020)\citenamefont
  {Blomberg}, \citenamefont {Rathnayake},\ and\ \citenamefont
  {Moreau}}]{Blomberg:2020}%
  \BibitemOpen
  \bibfield  {author} {\bibinfo {author} {\bibfnamefont {S.~P.}\ \bibnamefont
  {Blomberg}}, \bibinfo {author} {\bibfnamefont {S.~I.}\ \bibnamefont
  {Rathnayake}},\ and\ \bibinfo {author} {\bibfnamefont {C.~M.}\ \bibnamefont
  {Moreau}},\ }\bibfield  {title} {\bibinfo {title} {Beyond {B}rownian motion
  and the {O}rnstein-{U}hlenbeck process: stochastic diffusion models for the
  evolution of quantitative characters},\ }\href@noop {} {\bibfield  {journal}
  {\bibinfo  {journal} {Am. Nat.}\ }\textbf {\bibinfo {volume} {195}},\
  \bibinfo {pages} {145} (\bibinfo {year} {2020})}\BibitemShut {NoStop}%
\bibitem [{\citenamefont {Devi}\ and\ \citenamefont {Jain}(2023)}]{Devi:2023}%
  \BibitemOpen
  \bibfield  {author} {\bibinfo {author} {\bibfnamefont {A.}~\bibnamefont
  {Devi}}\ and\ \bibinfo {author} {\bibfnamefont {K.}~\bibnamefont {Jain}},\
  }\bibfield  {title} {\bibinfo {title} {Polygenic adaptation dynamics in
  large, finite populations},\ }\href@noop {} {\bibfield  {journal} {\bibinfo
  {journal} {bioRxiv:2023.01.25.525607}\ }\textbf {\bibinfo {volume} {-}},\
  (\bibinfo {year} {2023})}\BibitemShut {NoStop}%
\bibitem [{\citenamefont {Kimura}(1964)}]{Kimura:1964}%
  \BibitemOpen
  \bibfield  {author} {\bibinfo {author} {\bibfnamefont {M.}~\bibnamefont
  {Kimura}},\ }\bibfield  {title} {\bibinfo {title} {Diffusion models in
  population genetics},\ }\href@noop {} {\bibfield  {journal} {\bibinfo
  {journal} {J. Appl. Prob.}\ }\textbf {\bibinfo {volume} {1}},\ \bibinfo
  {pages} {177} (\bibinfo {year} {1964})}\BibitemShut {NoStop}%
\bibitem [{\citenamefont {Ewens}(2004)}]{Ewens:2004}%
  \BibitemOpen
  \bibfield  {author} {\bibinfo {author} {\bibfnamefont {W.}~\bibnamefont
  {Ewens}},\ }\href@noop {} {\emph {\bibinfo {title} {Mathematical Population
  Genetics}}}\ (\bibinfo  {publisher} {Springer, Berlin},\ \bibinfo {year}
  {2004})\BibitemShut {NoStop}%
\bibitem [{\citenamefont {Risken}(1996)}]{Risken:1996}%
  \BibitemOpen
  \bibfield  {author} {\bibinfo {author} {\bibfnamefont {H.}~\bibnamefont
  {Risken}},\ }\href@noop {} {\emph {\bibinfo {title} {The Fokker Planck
  equation. Methods of solution and applications}}}\ (\bibinfo  {publisher}
  {Springer, Berlin},\ \bibinfo {year} {1996})\BibitemShut {NoStop}%
\bibitem [{\citenamefont {McKane}\ and\ \citenamefont
  {Waxman}(2007)}]{McKane:2007}%
  \BibitemOpen
  \bibfield  {author} {\bibinfo {author} {\bibfnamefont {A.~J.}\ \bibnamefont
  {McKane}}\ and\ \bibinfo {author} {\bibfnamefont {D.}~\bibnamefont
  {Waxman}},\ }\bibfield  {title} {\bibinfo {title} {Singular solutions of the
  diffusion equation of population genetics},\ }\href@noop {} {\bibfield
  {journal} {\bibinfo  {journal} {J Theor Biol}\ }\textbf {\bibinfo {volume}
  {247}},\ \bibinfo {pages} {849} (\bibinfo {year} {2007})}\BibitemShut
  {NoStop}%
\bibitem [{\citenamefont {Olver}\ \emph {et~al.}(2024)\citenamefont {Olver},
  \citenamefont {{Olde Daalhuis}}, \citenamefont {Lozier}, \citenamefont
  {Schneider}, \citenamefont {Boisvert}, \citenamefont {Clark}, \citenamefont
  {Miller}, \citenamefont {Saunders}, \citenamefont {Cohl},\ and\ \citenamefont
  {McClain}}]{DLMF}%
  \BibitemOpen
  \bibfield  {author} {\bibinfo {author} {\bibfnamefont {F.~W.~J.}\
  \bibnamefont {Olver}}, \bibinfo {author} {\bibfnamefont {A.~B.}\ \bibnamefont
  {{Olde Daalhuis}}}, \bibinfo {author} {\bibfnamefont {D.~W.}\ \bibnamefont
  {Lozier}}, \bibinfo {author} {\bibfnamefont {B.~I.}\ \bibnamefont
  {Schneider}}, \bibinfo {author} {\bibfnamefont {R.~F.}\ \bibnamefont
  {Boisvert}}, \bibinfo {author} {\bibfnamefont {C.~W.}\ \bibnamefont {Clark}},
  \bibinfo {author} {\bibfnamefont {B.~R.}\ \bibnamefont {Miller}}, \bibinfo
  {author} {\bibfnamefont {B.~V.}\ \bibnamefont {Saunders}}, \bibinfo {author}
  {\bibfnamefont {H.~S.}\ \bibnamefont {Cohl}},\ and\ \bibinfo {author}
  {\bibfnamefont {M.~A.}\ \bibnamefont {McClain}},\ }\href@noop {} {\emph
  {\bibinfo {title} {NIST Digital Library of Mathematical Functions}}}\
  (\bibinfo  {publisher} {http://dlmf.nist.gov/, Release 1.2.3 of 2024-12-15},\
  \bibinfo {year} {2024})\BibitemShut {NoStop}%
\bibitem [{\citenamefont {Kimura}(1955)}]{Kimura:1955a}%
  \BibitemOpen
  \bibfield  {author} {\bibinfo {author} {\bibfnamefont {M.}~\bibnamefont
  {Kimura}},\ }\bibfield  {title} {\bibinfo {title} {Solution of a process of
  random genetic drift with a continous model},\ }\href@noop {} {\bibfield
  {journal} {\bibinfo  {journal} {Proc. Natl. Acad. Sci. USA}\ }\textbf
  {\bibinfo {volume} {41}},\ \bibinfo {pages} {144} (\bibinfo {year}
  {1955})}\BibitemShut {NoStop}%
\bibitem [{\citenamefont {Kubo}(1962)}]{Kubo:1962}%
  \BibitemOpen
  \bibfield  {author} {\bibinfo {author} {\bibfnamefont {R.}~\bibnamefont
  {Kubo}},\ }\bibfield  {title} {\bibinfo {title} {Generalized cumulant
  expansion method},\ }\href@noop {} {\bibfield  {journal} {\bibinfo  {journal}
  {J. Phys. Soc. Japan}\ }\textbf {\bibinfo {volume} {17}},\ \bibinfo {pages}
  {1100} (\bibinfo {year} {1962})}\BibitemShut {NoStop}%
\bibitem [{\citenamefont {Waples}(2022)}]{Waples:2022}%
  \BibitemOpen
  \bibfield  {author} {\bibinfo {author} {\bibfnamefont {R.~S.}\ \bibnamefont
  {Waples}},\ }\bibfield  {title} {\bibinfo {title} {What is {Ne}, anyway?},\
  }\href@noop {} {\bibfield  {journal} {\bibinfo  {journal} {Journal of
  Heredity}\ }\textbf {\bibinfo {volume} {113}},\ \bibinfo {pages} {371}
  (\bibinfo {year} {2022})}\BibitemShut {NoStop}%
\bibitem [{\citenamefont {Eriksson}\ \emph {et~al.}(2010)\citenamefont
  {Eriksson}, \citenamefont {Mehlig}, \citenamefont {Rafajlovic},\ and\
  \citenamefont {Sagitov}}]{Eriksson:2010}%
  \BibitemOpen
  \bibfield  {author} {\bibinfo {author} {\bibfnamefont {A.}~\bibnamefont
  {Eriksson}}, \bibinfo {author} {\bibfnamefont {B.}~\bibnamefont {Mehlig}},
  \bibinfo {author} {\bibfnamefont {M.}~\bibnamefont {Rafajlovic}},\ and\
  \bibinfo {author} {\bibfnamefont {S.}~\bibnamefont {Sagitov}},\ }\bibfield
  {title} {\bibinfo {title} {The total branch length of sample genealogies in
  populations of variable size},\ }\href@noop {} {\bibfield  {journal}
  {\bibinfo  {journal} {Genetics}\ }\textbf {\bibinfo {volume} {186}},\
  \bibinfo {pages} {601–611} (\bibinfo {year} {2010})}\BibitemShut {NoStop}%
\bibitem [{\citenamefont {Kimura}\ and\ \citenamefont
  {Ohta}(1969)}]{Kimura:1969}%
  \BibitemOpen
  \bibfield  {author} {\bibinfo {author} {\bibfnamefont {M.}~\bibnamefont
  {Kimura}}\ and\ \bibinfo {author} {\bibfnamefont {T.}~\bibnamefont {Ohta}},\
  }\bibfield  {title} {\bibinfo {title} {The average number of generations
  until fixation of a mutant gene in a finite population},\ }\href@noop {}
  {\bibfield  {journal} {\bibinfo  {journal} {Genetics}\ }\textbf {\bibinfo
  {volume} {61}},\ \bibinfo {pages} {763} (\bibinfo {year} {1969})}\BibitemShut
  {NoStop}%
\end{thebibliography}

%


\end{document}